\documentclass[aps,pra,twocolumn,superscriptaddress,amsmath,amssymb,showpacs]{revtex4-1}

\usepackage{graphicx}	% Include figure files
\usepackage{dcolumn}	% Align table columns on decimal point
\usepackage{bm}		% Bold math
\usepackage{verbatim} 	% Long comments
\usepackage{braket}
\usepackage{xr}
\usepackage{hyperref}

\begin{document}

\title{Rate-distance tradeoff and resource costs for all-optical quantum repeaters}


\author{Mihir Pant}
\email{mpant@mit.edu}
\affiliation{Dept. of Electrical Engineering and Computer Science, MIT, Cambridge, MA 02139, USA}
\affiliation{Quantum Information Processing group, Raytheon BBN Technologies, 10 Moulton Street, Cambridge, MA 02138, USA}

\author{Hari Krovi}
\affiliation{Quantum Information Processing group, Raytheon BBN Technologies, 10 Moulton Street, Cambridge, MA 02138, USA}

\author{Dirk Englund}
\affiliation{Dept. of Electrical Engineering and Computer Science, MIT, Cambridge, MA 02139, USA}

\author{Saikat Guha}
\affiliation{Quantum Information Processing group, Raytheon BBN Technologies, 10 Moulton Street, Cambridge, MA 02138, USA}

\begin{abstract}
We present a resource-performance tradeoff of an all-optical quantum repeater that uses photon sources, linear optics, photon detectors and classical feedforward at each repeater node, but no quantum memories. We show that the quantum-secure key rate has the form $R(\eta) = D\eta^s$ bits per mode, where $\eta$ is the end-to-end channel's transmissivity, and the constants $D$ and $s$ are functions of various device inefficiencies and the resource constraint, such as the number of available photon sources at each repeater node. Even with lossy devices, we show that it is possible to attain $s < 1$, and in turn outperform the maximum key rate attainable without quantum repeaters, $R_{\rm direct}(\eta) = -\log_2(1-\eta) \approx (1/\ln 2)\eta$ bits per mode for $\eta \ll 1$, beyond a certain total range $L$, where $\eta \sim e^{-\alpha L}$ in optical fiber. We also propose a suite of modifications to a recently-proposed all-optical repeater protocol that ours builds upon, which lower the number of photon sources required to create photonic clusters at the repeaters so as to outperform $R_{\rm direct}(\eta)$, from $\sim 10^{11}$ to $\sim 10^{6}$ photon sources per repeater node. We show that the optimum separation between repeater nodes is independent of the total range $L$, and is around $1.5$ km for assumptions we make on various device losses.
\end{abstract}

\maketitle

\section{Introduction}\label{sec:intro}
Quantum key distribution (QKD) enables two distant authenticated parties Alice and Bob, connected via a quantum (e.g., optical) channel, to generate information-theoretically secure shared secret bits. No knowledge of the channel conditions (noise model, or any channel estimate) is required a priori to ensure security. However, the shared secret is generated at a rate commensurate with the worst-case adversary physically consistent with the channel conditions actually presented to Alice and Bob. The reason is that all the perceived channel imperfections (absolutely anything that causes the channel map to deviate from a noiseless identity transformation) is attributed to the actions of the most powerful adversary allowed by physics---even though some (or all) of that deviation of the channel from an identity map may actually stem from non-adversarial sources, such as losses due to free-space diffraction, fiber loss, detection inefficiency, thermal noise from blackbody at the operating temperature and wavelength, and detector noise. An important consequence of this assumption is that all the signal power transmitted by Alice that is not collected by Bob is made available coherently to the eavesdropper, Eve. This model for Eve is the intuition behind why the secret key rate for a direct-transmission based QKD protocol must decrease linearly with $\eta$, the Alice-Bob power transmissivity, in the $\eta \ll 1$ regime~\cite{2014.NatureComm.Takeoka-Wilde.TGWBound, 2015.ArXiv.Pirandola-Banchi.QuantCommUltRate}. For any direct-transmission protocol over the pure-loss optical channel of transmissivity $\eta$, and assuming unlimited authenticated two-way public classical communication, it was recently shown that the key rate cannot exceed $-\log_2(1-\eta)$ bits per mode~\cite{2015.ArXiv.Pirandola-Banchi.QuantCommUltRate}, which is $\approx 1.44\eta$ for $\eta \ll 1$. For a pure-loss channel, the Pirandola-Laurenza-Ottaviani-Banchi (PLOB) upper bound improves over the Takeoka-Guha-Wilde (TGW) bound~\cite{2014.NatureComm.Takeoka-Wilde.TGWBound} by a factor of $2$ in the $\eta \ll 1$ regime. The TGW bound is an upper bound on the secret-key agreement capacity with unlimited two-way classical communication $P_2({\cal N})$, applicable to a general quantum channel ${\cal N}$. For the pure-loss channel ${\cal N}_\eta$, the PLOB bound coincides with the best-known achievable rate~\cite{2009.PRL.Pirandola-Lloyd.SKCLB}, thus establishing $P_2({\cal N}_\eta) = -\log_2(1-\eta)$ bits per mode. From hereon, we denote by $R_{\rm direct}(\eta) \equiv -\log_2(1-\eta)$ the maximum bits-per-mode secret key rate achievable by any direct-transmission QKD protocol, i.e., without the use of quantum repeaters. The bits/s rate of a QKD protocol's implementation is obtained by multiplying the bits/mode rate by the spatio-temporal-polarization bandwidth (modes/s), which is governed by the channel geometry, and the transmitter and detector bandwidth. Since loss increases exponentially with distance $L$ in optical fiber (i.e., $\eta = e^{-\alpha L}$), for $\eta \ll 1$, the key rate generated by any direct-transmission QKD protocol must decay exponentially with the range $L$. Expressed as a function of $L$, $R_{\rm direct}(L) = -\log_2(1-e^{-\alpha L})$ bits/mode, which is $\approx 1.44 e^{-\alpha L}$ bits/mode, for $L$ large.

Quantum repeaters, proposed in ~\cite{1998.PRL.Briegel-Zoller.BDCZ}, are devices which when inserted along the length of the optical channel, can help generate shared secret at a rate that surpasses $R_{\rm direct}(\eta)$ at any value of Alice-to-Bob channel transmissivity $\eta$~\cite{2015.ArXiv.Pirandola-Banchi.QuantCommUltRate}. Quantum repeaters need not be trusted or physically secured in order to ensure the security of the keys generated. If $n$ quantum repeaters are inserted along the length of the channel connecting the communicating parties Alice and Bob, and if there are absolutely no physical constraints placed on the repeater nodes (i.e., the repeaters are assumed to be lossless, error-corrected, general purpose quantum computers), then the maximum key rate achievable by Alice and Bob is given by $-\log_2(1-\eta_{\rm min})$ bits/mode, where $\eta_{\rm min} \equiv {\rm {min}}\left(\eta_1, \eta_2, \ldots, \eta_{n+1}\right)$, with $\eta = \eta_1\ldots\eta_{n+1}$, is the transmissivity of the lossiest link between successive repeater nodes~\cite{2016.ArXiv.Pirandola-P2P_Network} (see~\cite{2016.ArXiv.Azuma-P2P_Network} for a different upper bound based on squashed entanglement~\cite{2014.NatureComm.Takeoka-Wilde.TGWBound}). Given $n$ ideal repeater nodes, their optimal placement is to lay them equally-spaced, in which case, the maximum achievable rate is $-\log_2(1-\eta^{1/(n+1)})$ bits/mode. As $n \to \infty$, the rate is unbounded. However, assuming repeaters to be lossless error-corrected quantum computers is not practical. A more practically relevant question to ask is if the repeater nodes have finite resources with lossy and imperfect components (where `resources' may be different physical entities depending upon the type of quantum repeater and the protocol employed), then what rate can Alice and Bob achieve, and more importantly what would it take to build repeater nodes so as to be able to significantly outperform $R_{\rm direct}(\eta) = -\log_2(1-\eta)$ bits/mode. This is the topic addressed in this paper, for repeaters that are built solely using photonic components---single-photon sources, detectors, electro-optic feedforward, but no matter-based quantum memories. As we will see later in this paper, that given physical constraints on a repeater node, placing more repeaters (higher $n$) between Alice and Bob may not always improve the rate, i.e., depending upon the total distance $L$ (or equivalently, the transmissivity $\eta$) between Alice and Bob, and given the physical device constraints in a repeater node, there may be an optimal number $n^*(\eta)$ of nodes, which achieves the highest end-to-end rate.

Traditional quantum repeaters work in the framework of entanglement-based QKD. At the end of the transmission phase of an entanglement-based QKD protocol, Alice and Bob share (noisy, or imperfect) entangled pairs (e.g., of photons or matter-based stationary qubits) which could have been tampered with, or which could have deteriorated due to channel loss and noise. At that point, if the end goal of Alice and Bob is to generate shared entanglement (for use in some quantum protocol that consumes shared entanglement, such as teleporation~\cite{1993.PRL.Bennett-Wootters.Teleportation} or dense coding~\cite{1992.PRL.Bennett-Wiesner.SuperdenseCoding}), they would perform {\em entanglement distillation} to sieve out a small number of clean maximally entangled Bell pairs by performing local operations and classical communications (LOCC). If the end goal of Alice and Bob is to generate shared secret (a strictly less demanding goal than generating shared entanglement), they directly measure the noisy shared entangled pairs, and perform (classical) error correction and privacy amplification on their correlated measurement results over an authenticated public channel to distill a quantum-secure shared secret key. 

Several different genres of repeater protocols have been proposed~\cite{2015.ArXiv.Muralidharan-Jiang.RepeaterGen}. The two primary ingredients in any of the traditional repeater architectures are: (1) some form of a quantum memory, and (2) the ability to perform a certain restricted class of quantum logic, i.e., gates and measurements on the flying (photonic) qubits as well as the static (memory) qubits. In the most basic repeater protocol, the restricted quantum operation required is Bell state measurement (BSM) on pairs of qubits. A BSM on qubit $b$ and qubit $c$ converts two independent Bell pairs $|\Psi\rangle^{ab}$ and $|\Psi\rangle^{cd}$ into one Bell pair $|\Psi\rangle^{ad}$, upto local single qubit operations, a process known as {\em entanglement swapping}.

\subsection{Quantum repeaters based on mode multiplexing and Bell state measurements}

In the following discussion, we will focus on a class of quantum repeaters that rely solely on probabilistic BSMs, quantum memories, and multiplexing, i.e., the ability to `switch' qubits across (spatial, spectral, or temporal) modes. The essence of such a repeater protocol was developed by Sinclair {\em et al.}~\cite{2014.PRL.Sinclair-Tittel.AtomFreqCombRep}, which employed spectral multiplexing in multimode quantum memories across $m$ parallel (spectral) channels, and entanglement swapping using linear optics and single photon detectors (the success probability of which can at most be $50\%$). Guha {\em et al.} analyzed the secret key rates achievable by the above protocol, with a fixed $m$ (memory size) and found that even when photon loss is the only source of noise, the achievable key rate is of the form $R(\eta) = D\eta^s$ bits/mode, where $D$, and $s < 1$ are constants that are functions of various losses in the system (e.g., detection efficiencies, coupling losses, memory loading and readout efficiencies, and BSM failure probability)~\cite{2015.PRA.Guha-Tittel.QRRateLossAnalysis}. Since the exponent of $\eta$, i.e., $s$ is strictly less than $1$, the key rate must beat $R_{\rm direct}(\eta)$ (which scales as: $\propto \eta$ for $\eta \ll 1$) beyond a certain minimum distance determined by the actual values of the system's loss parameters, which is around a couple of hundred kilometers for reasonable estimates of the losses~\cite{2015.PRA.Guha-Tittel.QRRateLossAnalysis}. Since $\eta = e^{-\alpha L}$ in fiber, the rate achieved by this repeater protocol for a fixed memory size, $R(L) = De^{-s\alpha L}$ still scales exponentially with the range $L$, albeit with a smaller exponent compared to the best possible rate without any repeater, which could turn into a huge absolute improvement in the end-to-end secret key rate~\cite{2015.PRA.Guha-Tittel.QRRateLossAnalysis}. 

Azuma {\em et al.} recently proposed an all-photonic variant of this protocol in which they substituted matter based quantum memories with optical cluster states~\cite{2015.NatureComm.Azuma-Lo.AllOptRep}, based on a proposal by Varnava {\em et al.} to mimic a quantum memory (i.e., protect against photon losses) by appending each physical photonic qubit by an entangled `tree cluster' state~\cite{2007.NJP.Varnava-Rudolph.CountFactECMemory}. As long as the losses incurred by each photon (i.e., photons being protected as well as the additional photons in the trees added for loss protection) is less than $3$ dB, the effective loss of the logical qubit can be made to approach zero, by increasing the size of the tree cluster, i.e., the number of photons in the logical qubit~\cite{2006.PRL.Varnava-Rudolph.CountEC}. Thus, Azuma {\em et al.}'s proposal showed the theoretical feasibility of a quantum repeater architecture (i.e., one that can beat the scaling of direct-transmission QKD) using only flying qubits, with the repeater {\em nodes} being equipped only with single photon sources, passive linear-optical circuits (beamsplitters and phase shifters), single photon detectors, and classical feedforward.

Azuma {\em et al.}'s result marked a promising conceptual leap towards all-optical quantum repeaters. However, important unanswered questions remained, including the achievable secure key generation rate and how it scales with distance (or loss), as well as the physical resource requirements: e.g., the number of photon sources and detectors at the repeater nodes. As an example, a calculation in their paper shows that at a range of $L = 5000$ km, an entanglement-generation rate of $69$ kHz is achievable in a fiber based linear optic system with $100$ kHz repetition rate, $150$ ns feed forward time and a source-detector efficiency product of $95\%$  whereas sharing a single entangled photon pair via a direct transmission scheme with the same parameters would require $10^{81}$ years. The level of error protection required to achieve the aforesaid repeater performance at $L=5000$ km would require one to build entangled clusters of $\sim 10^4$ photons at the $100$ kHz clock rate at each repeater node. Building such a cluster using linear optics and feed-forward~\cite{2008.PRL.Varnava-Rudolph.PhotonLoss-LOQCscaling, 2015.PRX.Li-Benjamin.RescostFTLOQC} would require around $10^{24}$ single photon sources at each repeater node. Furthermore, since every photon used for error correction is sent between repeater nodes in ~\cite{2015.NatureComm.Azuma-Lo.AllOptRep}, their scheme would require around $20,000$ parallel channels connecting the neighboring nodes. Thus, while Ref.~\cite{2015.NatureComm.Azuma-Lo.AllOptRep} showed the theoretical possibility of all-optical repeaters, clearly further work is needed to address their practical feasibility. These results open up a compelling line of research to investigate improved all-photonic repeater architectures of various genres which could be built with practically feasible resources, and also a thorough comparative study of rates achievable with each such all-optical repeater scheme.

\subsection{Main results}

Our contributions in this paper are twofold. The first is a rigorous analysis of: (a) the secret key rates achievable with the the aforesaid all-photonic repeater architecture given the size of the clusters generated at each repeater station, and (b) the resources required (e.g., number of single photon sources and detectors required at each repeater node) to build that cluster, while taking into account in explicit detail each step in building the required clusters using a network of passive linear optics (i.e., beamsplitters and phase shifters), imperfect on-demand sources with loss (see section \ref{sec:prelim} for a description of the source), single photon detectors (with some number resolving capability), and feed-forward. We find that the achievable secret key rate scales as $D\eta^s$ bits/mode, where $D$ and $s < 1$ are functions of the number of photon sources at each repeater node (the resource constraint---which is parametrically related to the size of the cluster), all the `inline' losses (e.g., losses in the optical fiber or waveguide used while creating the cluster, independent of the fiber loss between repeater stations), and the source and detector efficiencies. With $\eta \sim e^{-\alpha L}$ in fiber, the key rate still scales exponentially with $L$, but with a smaller exponent compared to the best direct-transmission protocol. This is no surprise given the analysis of ~\cite{2015.PRA.Guha-Tittel.QRRateLossAnalysis}, since the tree-cluster construction of~\cite{2015.NatureComm.Azuma-Lo.AllOptRep} essentially mimics an imperfect quantum memory, but one whose efficiency cannot simply be modeled by a constant per mode as in Ref.~\cite{2015.PRA.Guha-Tittel.QRRateLossAnalysis}. Using the cluster building scheme proposed by Li et al.~\cite{2015.PRX.Li-Benjamin.RescostFTLOQC}, we find that to a good approximation, the resource requirements are determined by the number of probabilistic fusion steps $k$ required to build the cluster starting from single photons, and hence, we calculate the performance with the best cluster that can be built in $k$ fusion steps. We use the scheme of Li {\em et al.} because it has been shown to be more efficient than the scheme of Varnava {\em et al.}~\cite{2008.PRL.Varnava-Rudolph.PhotonLoss-LOQCscaling} at building clusters~\cite{2015.PRX.Li-Benjamin.RescostFTLOQC}. Given all the inline and device losses, we evaluate the number of photon sources (and detectors) needed at each repeater node to beat $R_{\rm direct}(L)$ at a given total range $L$ between Alice and Bob. We also prove that given the device losses, there is an optimal spacing between the repeater nodes (which evaluates to roughly $1.5$ km for a set of system parameters we choose), regardless of the overall range $L$.

Our second major contribution in this paper is a significant improvement to the all-photonic repeater architecture in ~\cite{2015.NatureComm.Azuma-Lo.AllOptRep}---both in terms of the resources required at each node and the number of parallel optical channels connecting the neighboring nodes. We find that barely beating $R_{\rm direct}(L)$ using the all-optical scheme of~\cite{2015.NatureComm.Azuma-Lo.AllOptRep} requires more than $10^{11}$ photon sources at each repeater node for realizing the required optical cluster states and measurements. It also requires $208$ parallel channels connecting neighboring nodes, even when assuming very optimistic device-loss parameters. Assuming the same device losses, our improved repeater architecture reduces the number of photon sources (to barely beat $R_{\rm direct}(L)$) by $5$ orders of magnitude, while reducing the number of parallel channels to $8$. In both of these calculations, each source is used only once per clock cycle, i.e., they are not temporally multiplexed. We prove a tight analytical lower bound for the performance of our improved scheme. These performance advances are enabled primarily by the following: (1) using boosted fusion logic that improves the success probability of the BSM to $75\%$ by using four ancilla single photons~\cite{2014.PRL.Ewert-Loock.boostedfusion}, (2) employing a more resource-efficient scheme for creating tree clusters, building on the work of~\cite{2008.PRL.Varnava-Rudolph.PhotonLoss-LOQCscaling,2015.PRX.Li-Benjamin.RescostFTLOQC}, (3) retaining all the ancilla photons used for loss protection (i.e., to mimic a quantum memory) locally at the repeater nodes in a lossy waveguide, and (4) optimizing the timing of several single qubit measurements in the entire protocol.

We will limit our analysis to include photon losses (during the entire `lifetime' of each photon, i.e., from the time of generation to detection) but will not consider `multi-photon' errors stemming, for instance, from multi-photon emissions from the source, or detector dark clicks. We should note however that the error correction scheme analyzed here also provides some protection against depolarizing noise~\cite{2015.NatureComm.Azuma-Lo.AllOptRep}, a variant of which arises when one assumes multi-photon errors, and errors stemming from imperfect mode matching within the passive linear optical circuits at the repeater nodes.

The remainder of the paper is organized as follows. Section~\ref{sec:prelim} reviews preliminaries and notation used in the paper. Section~\ref{sec:architecture} describes our (improved) all-photonic quantum repeater architecture with a detailed description of each step starting from the creation of the tree clusters for error-protection, photon transmission, measurements at the repeater nodes, and the measurements by Alice and Bob, followed finally by key generation. Section~\ref{sec:rates} derives a closed form expression for a lower bound to the rate-distance envelope (i.e., an achievable rate by the protocol), which we show (numerically) to match the true rate-distance envelope extremely closely. Section~\ref{sec:discussion} compares our scheme to that of Ref.~\cite{2015.NatureComm.Azuma-Lo.AllOptRep} in terms of resource requirements and rates, and discusses possible avenues for further improvement. The concluding section~\ref{sec:conclusions} provides concrete directions for future research in order to further improve the prospects of a quantum communications network based solely on flying qubits. 

\section{Preliminaries}\label{sec:prelim}
\begin{figure}[h]
    \includegraphics[width=\columnwidth]{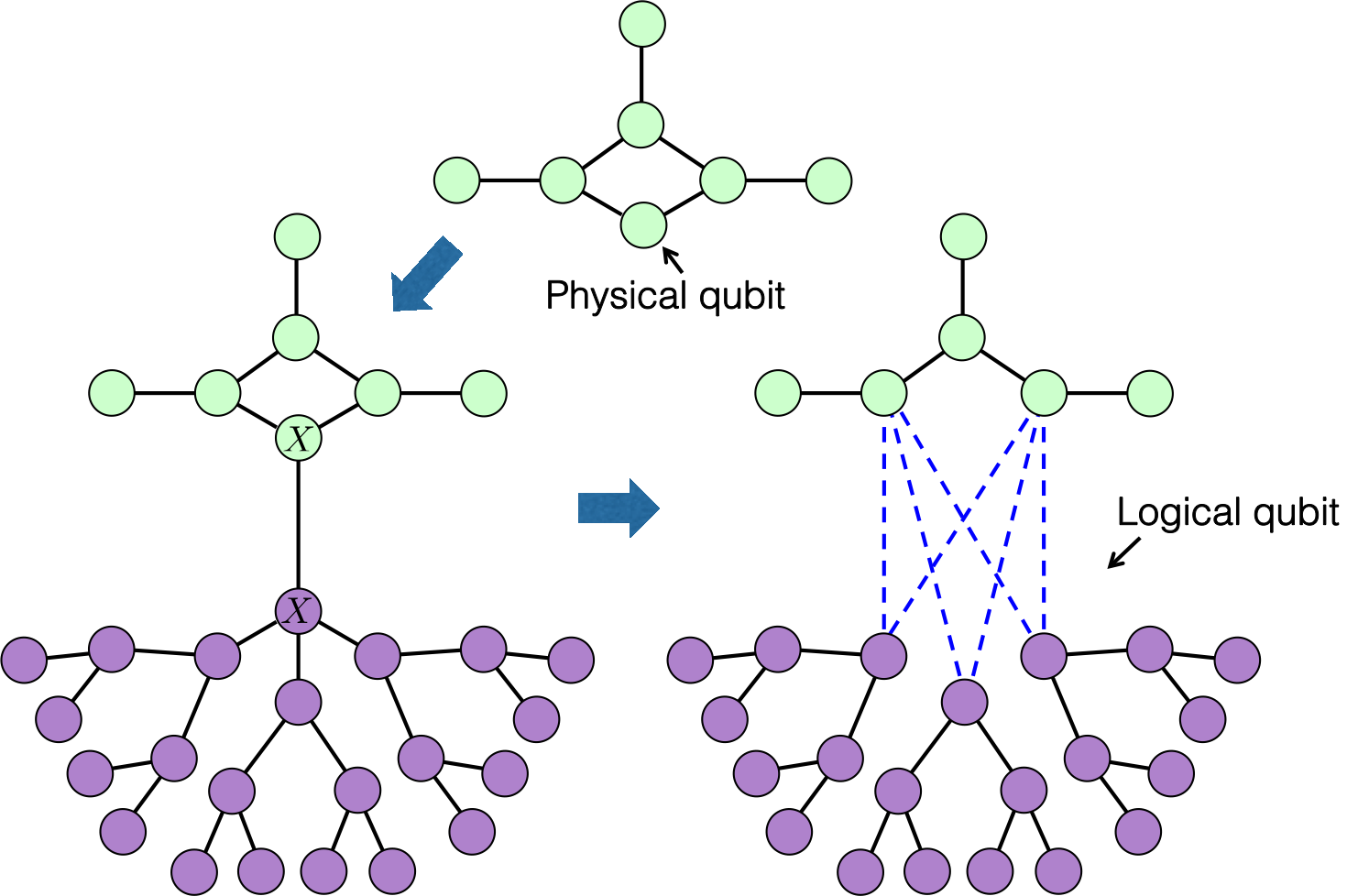}
    \caption{Attaching a $\left\{3,2,2\right\}$ tree to a node of a photonic cluster.} 
    \label{fig:attachtree}
\end{figure}

In this paper, we work with dual-rail photonic qubits, where the logical $|0\rangle$ and $|1\rangle$ are encoded by a single photon in one of two orthogonal (spatial) modes. A photonic {\em cluster} state (or, graph state), on a graph $G(V, E)$ with vertices in set $V$ and edges in set $E$, can be constructed by preparing each of the $|V|$ qubits (one stationed at each vertex) in the state $(\ket{0} + \ket{1})/\sqrt{2}$, and applying $|E|$ controlled-phase operations (a two-qubit unitary gate that applies a pauli $Z$ gate  to the second qubit if the first qubit is in the $\ket{1}$ state and applies an identity otherwise) on each pair of vertices that share an edge~\cite{2001.PRL.Raussendorf-Briegel.ClusterStateComp}. The (entangled) quantum state of the $|V|$ qubits thus obtained is an eigenstate of the $|V|$ stabilizer operators $X_i \; \Pi_{j \in {\cal N}(i)}Z_j$, where the index $i$ runs over all the vertices, $X_i$ and $Z_j$ are Pauli $X$ and $Z$ operators on qubit $i$ and qubit $j$ respectively, and ${\cal N}(i)$ is the set of all nearest neighbor vertices of vertex $i$. One simple observation, given that the cluster state is an eigenstate of the aforesaid stabilizer operators, is that an $X$ measurement on qubit $i$, and $Z$ measurements on all but one of the qubits in ${\cal N}(i)$, would deterministically reveal what the outcome of a $Z$ measurement on that unmeasured qubit in ${\cal N}(i)$ would have been, {\em even if} that unmeasured qubit had been lost. This realization is at the heart of the tree-based counterfactual error correction for protection against photon losses, developed by Varnava et al.~\cite{2006.PRL.Varnava-Rudolph.CountEC}. The idea is to attach a tree cluster to each physical photonic qubit in the graph state that needs to be protected against qubit loss. One can then deduce the result of any measurement on that qubit via an appropriate sequence of measurements on the qubits of the attached tree. The physical qubit and the qubits of the tree together form a protected (logical) qubit. We consider regular trees described by the {\em branching vector} ${\vec b} \equiv \left\{b_0, b_1, \ldots, b_m\right\}$, which signifies that the root of the tree has $b_0$ children nodes, and each of those nodes have $b_1$ children nodes, and so on until $b_0b_1\ldots b_m$ nodes at depth $m$. For such regular trees used for loss-error protection, one can write an explicit, yet recursive, expression for the success probability $P$ of performing an arbitrary single-qubit measurement on the protected qubit~\cite{2006.PRL.Varnava-Rudolph.CountEC}. It was shown that one can push $P$ arbitrarily close to $1$ as long as the probability of losing each photon is less than $1/2$. Fig.~\ref{fig:attachtree} illustrates how to attach a $\left\{3,2,2\right\}$ tree, shown by the dark (purple) shaded nodes, to a physical qubit of a cluster, shown by light (green) shaded nodes. Note that after the tree cluster is attached to the physical qubit, $X$ basis measurements must be performed on the physical qubit itself and the root node of the tree. These $X$ basis measurements, if successful, create additional edges (shown in dashed blue in Fig.~\ref{fig:attachtree}) between each neighboring qubit of the root node and each neighboring qubit of the physical qubit, after which the tree-protected logical qubit is ready to use. 

An ideal loss-less photonic cluster state on graph $G$ is a pure state, $\ket{\psi}_G$. A lossy cluster state on $G$ is obtained when all the photonic qubits of $\ket{\psi}_G$ are transmitted through independent pure-loss beamsplitters each of transmissivity $\eta$.  We call $1-\eta$ the {\em loss rate} of such a lossy cluster state. Clearly, the loss rate of $\ket{\psi}_G$ itself is $0$.

Arbitrary photonic cluster states can be prepared---with non-unity probability---using ideal single photons, passive linear optics (i.e., beamsplitters and phase shifters) and single photon detectors~\cite{2005.PRL.Browne-Rudolph.Clusterstates}. As examples, in the absence of losses, a two-photon maximally entangled (Bell) state can be prepared with success probability $3/16$~\cite{2008.PRA.Zhang-Pan.HeraldedBellPairsource}, whereas a three-photon maximally entangled (GHZ) state can be prepared with success probability $1/32$~\cite{2008.PRL.Varnava-Rudolph.PhotonLoss-LOQCscaling}. Browne and Rudolph introduced linear-optical Type I and Type II two-qubit fusion gates, which if successful (with probability $1/2$), can fuse two cluster fragments into one, according to specific rules~\cite{2005.PRL.Browne-Rudolph.Clusterstates}. These fusion gates, in conjunction with Bell states and GHZ states, can be used to construct arbitrary cluster states. The success probability of the fusion gates can be improved to $3/4$ if additional (ancilla) single photons are available to be injected on-demand into an otherwise-passive linear optical circuit, and if the detectors have up to two-photon number resolution~\cite{2014.PRL.Ewert-Loock.boostedfusion}. We assume such {\em boosted} fusion gates in our all-optical repeater construction described in this paper.

We model a lossy single photon source of efficiency $\eta_s$ as one that emits, on demand, the mixed state $\eta_s\ket{1}\bra{1}+(1-\eta_s)\ket{0}\bra{0}$. We use $\eta_d$ for the efficiency of all detectors in the system. We will assume that the cluster is created on a photonic chip to allow for easier scalability after which the photons are coupled, with efficiency $\eta_c$, into fiber with loss coefficient $\alpha$ and speed of light $c_f$. $P_{\rm chip} = e^{-\beta \tau_s c_{\rm ch}}$ denotes the survival probability of a photon on-chip during one feed-forward step, where $\beta$ is the loss coefficient, $c_{\rm ch}$ is the speed of light and $\tau_s$ is the feed-forward time, all on-chip. ${\eta}_{\rm GHZ} = \eta_s\eta_d/(2-\eta_s\eta_d)$ is the survival rate of the photons that are input into a linear-optical circuit intended to produce $3$-photon maximally-entangled GHZ states~\cite{2008.PRL.Varnava-Rudolph.PhotonLoss-LOQCscaling}. The final measurement step requires feed-forward in fiber. The survival probability, $P_{\rm fib}$, during feed-forward time in fiber, $\tau_f$, is $P_{\rm fib} = e^{-\alpha \tau_f c_{\rm f}}$. The values for device performance assumed for the plots that appear later in the paper, are summarized in Table~\ref{deviceparamtable}.

\section{Repeater Architecture}\label{sec:architecture}
\begin{figure}[h]
    \includegraphics[width=0.5\textwidth]{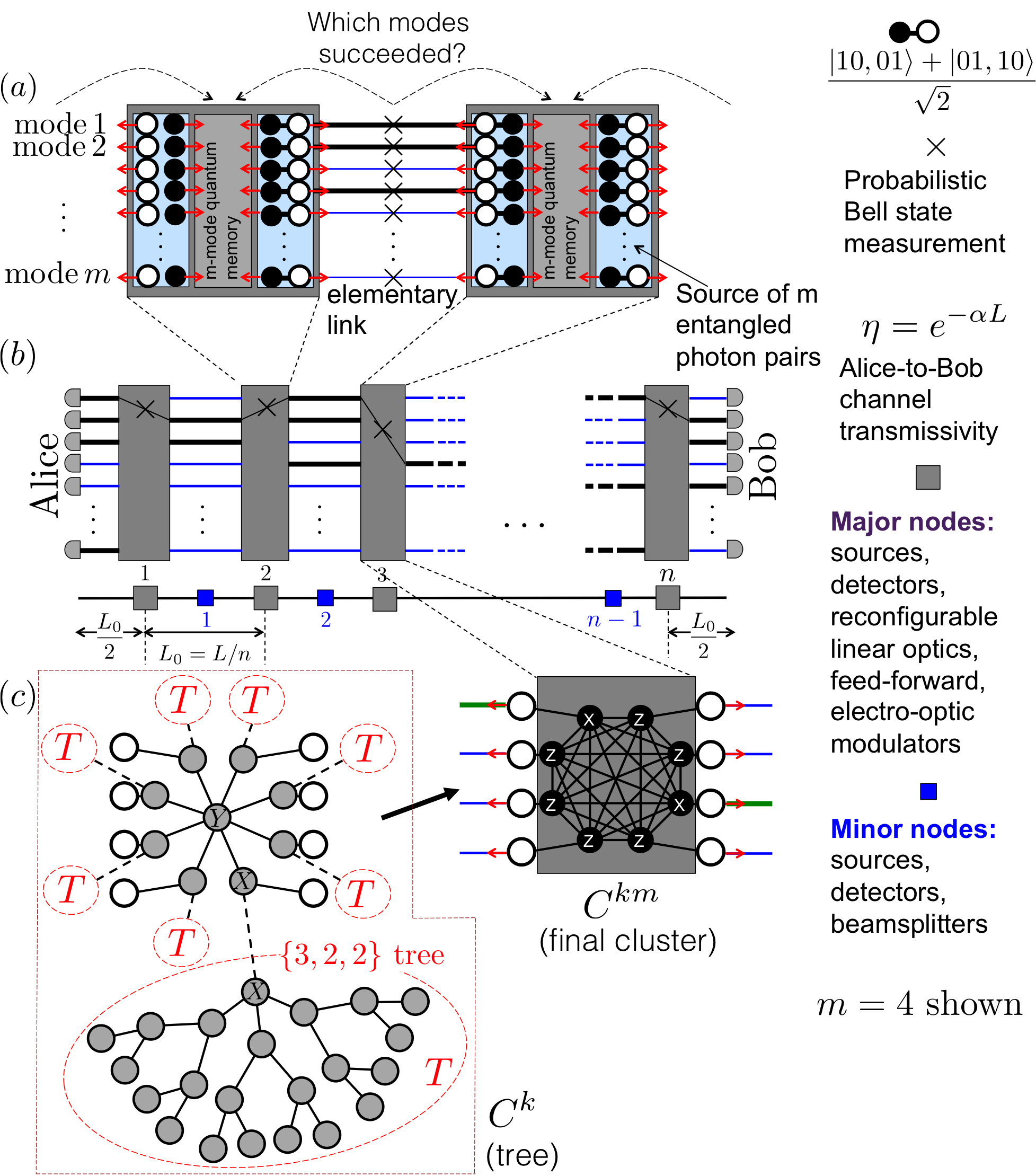}
    \caption{(a) and (b) show schematics of one elementary link, and a chain of them connecting Alice and Bob, respectively, for a repeater architecture that employs quantum memories, Bell pair sources, probabilistic BSMs, and multiplexing over $m$ orthogonal parallel channels. (c) depicts the construction of a photonic cluster state that can subsume the roles of the quantum memory and the Bell pair sources, thereby resulting in a quantum repeater architecture based solely on `flying' qubits. The outer (white) photonic qubits are transmitted on the fiber channels, and the inner (black) qubits are held locally in a (lossy) waveguide at the repeater node. See text for a detailed description.} 
    \label{Lorepeaterconcept}
\end{figure}

Before we discuss the all-photonic repeater architecture, it is instructive to review a generic quantum repeater architecture based on multimode quantum memories, probabilistic BSMs, and multiplexing over $m$ parallel channels depicted in Figs.~\ref{Lorepeaterconcept}(a) and (b), which was proposed by~\cite{2014.PRL.Sinclair-Tittel.AtomFreqCombRep}, and analyzed in~\cite{2015.PRA.Guha-Tittel.QRRateLossAnalysis}. The parallel channels can be a combination of mutually-orthogonal spectral, spatial, and polarization modes, over each of which dual-rail photonic qubits can be transmitted simultaneously at the clock rate (determined by the source and detector bandwidth). Alice and Bob are separated by optical fiber of length $L$ (i.e., end-to-end transmissivity, $\eta = e^{-\alpha L}$), interspersed with $n$ repeater stations spaced $L_0 = L/n$ apart, with Alice and Bob $L_0/2$ away from the terminal repeaters in the chain. 

Each of the $n$ repeater nodes (or, `major nodes'), shown by a gray box, consists of a multimode quantum memory straddled between sources of $m$ Bell pairs on its left and another $m$ on its right. Each major node loads one half of an entangled Bell state onto the memory, while transmitting the other half towards the middle of the adjoining {\em elementary link}. Each major node does the above synchronously on every clock cycle. At the center of each elementary link is a `minor node', shown as dark-blue-shaded boxes in Fig.~\ref{Lorepeaterconcept}(b). After the qubits from the major nodes reach the minor nodes (i.e., after propagation through a distance $L_0/2$), each minor node, simultaneously, performs BSMs on each of the $m$ pairs of qubits received from the repeater nodes on its either side. The successful BSMs within each elementary link are shown by thick (green) line segments. Immediately after the minor node BSMs, each minor node sends back the information---about which of the $m$ channels were successfully measured---to its two neighboring major nodes, on an authenticated classical channel. Upon receipt of that information, each major node performs a BSM on two qubits held in its memory that had been entangled halves of qubits that participated in successful BSMs at the minor node to the left of that major node, and the minor node to its right, respectively. Simultaneous with the minor-node BSMs, Alice and Bob measure, in one of the two randomly-chosen mutually-unbiased bases, the $m$ qubits they receive at their respective ends of the terminal half-elementary-link segments (see Fig.~\ref{Lorepeaterconcept}(b)), and send the information about which channels generated a `click' on their detectors, back to their respective neighboring major nodes. Finally, each major node sends the information on whether its BSM succeeded, to Alice and Bob. Hence, at every clock cycle, with some probability (i.e., if all the minor nodes heralded at least one success each, all major node BSMs were successful, and Alice and Bob both detected a photon on at least one of the $m$ channels each while using the same measurement bases), Alice and Bob obtain a shared (raw, sifted) bit. A long sequence of sifted bits is thereafter used to distill a quantum-secure shared secret via error correction and privacy amplification.

The all-optical repeater architecture we now discuss builds upon a recent proposal by Azuma {\em et al.}~\cite{2015.NatureComm.Azuma-Lo.AllOptRep}, although there are some important differences, which we will point out later in Section~\ref{sec:discussion}. The key idea is to mimic a quantum memory (whose goal is essentially to protect photonic qubits against loss for a certain time duration) by using the tree cluster approach described in Section~\ref{sec:prelim}. The authors of~\cite{2015.NatureComm.Azuma-Lo.AllOptRep} went one step further and subsumed the functionalities of all the subcomponents of the major node (the quantum memory as well as the $2m$ Bell pair sources) into one single giant optical cluster state, which we describe next. Fig.~\ref{Lorepeaterconcept}(c) illustrates the construction of this cluster. We start with a depth-$2$ star cluster with a degree-$2m$ root node, and $4m+1$ total qubits. The `outer' qubits, shown as white circles, play a role analogous to the white qubits in Fig.~\ref{Lorepeaterconcept}(a) that are transmitted to the minor nodes on fiber channels. The $2m$ `inner' qubits, shown as gray circles, are each attached with a tree cluster of an appropriately-chosen branching vector ${\vec b}$, thereby creating a giant tree cluster. The loss-protected (logical) inner qubits play a dual role, that of the black qubits in Fig.~\ref{Lorepeaterconcept}(a) that are held in the quantum memories locally at the major nodes, and that of the memories themselves. We make the two $X$ measurements corresponding to each tree appended to the star, as described in the previous section (i.e., a total of $4m$ $X$ measurements). Finally, we make a $Y$ measurement on the root node of the star, which has an effect of creating a clique among all the (logical) inner qubits, shown by black circles in Fig.~\ref{Lorepeaterconcept}(c). The clique of the $2m$ logical inner qubits, connected to the $2m$ outer qubits, forms the full photonic cluster state that each major node creates every clock cycle, and sends out the $2m$ outer qubits (the white circles) towards the neighboring minor nodes ($m$ to the left and $m$ to the right) on fiber channels. Note that the final cluster state (after the $X$ and $Y$ measurements) is not a tree.

Each major node is equipped with single photon sources, reconfigurable passive linear optics, and single photon detectors. The clusters are created using linear optics and feed-forward~\cite{2008.PRL.Varnava-Rudolph.PhotonLoss-LOQCscaling, 2015.PRX.Li-Benjamin.RescostFTLOQC}. Since the cluster creation process is probabilistic, the resources (number of photon sources, detectors, size of linear optic circuit) must be chosen to ensure a near-unity success probability of creating the cluster in every clock cycle (see Fig.~\ref{PcnvsNs}). 

The minor nodes are identical to what was described earlier. The remainder of the protocol proceeds exactly as described at the beginning of this Section in the context of the memory-based architecture, except for the following difference of the action at the major nodes. When the information about which modes were successful comes back at a major node (from the two neighboring minor nodes), instead of doing a BSM between a pair of qubits held in a memory, the major node applies $X$ measurements on the two logical inner qubits corresponding to the successful modes on either side of the clique, and makes $Z$ measurements on the remaining $2m-2$ logical inner qubits (see Fig.~\ref{Lorepeaterconcept}(c)). The $X$ measurements have the effect of fusing the successful outer qubits into an entangled chain, and the $Z$ measurements have the effect of removing the extraneous qubits from the cluster. 

So, in any given clock cycle, if the photonic clusters at each major node are successfully created (which includes success in performing the $4m$ $X$ measurements and one $Y$ measurement), if all the minor nodes herald at least one BSM success, if the logical (inner) qubits survive the local storage at the major nodes while the outer qubits fly to the minor nodes and the classical information (about which modes were successful) arrives back, if the two $X$ measurements and $2m-2$ $Z$ measurements done to prune the clusters at the major nodes using that classical information are successful, and if Alice and Bob get at least one click each while using same measurement bases, then Alicte and Bob obtain a raw sifted shared bit. In Section~\ref{sec:rates}, we explicitly calculate this overall success probability, and the resulting secret-key generation rate. As we will see, larger error-protection trees afford better rate performance (up to a limit governed by the device loss rates), but creating larger clusters at the major nodes requires more resources (sources and detectors).

In Section~\ref{sec:creatingclique}, we describe in detail the construction of the clusters at the major nodes using linear optics, and calculate the success probability. In Section~\ref{sec:measuringclusters}, we will describe how the measurements on the major-node clusters are done, after the BSMs at the minor nodes, to stitch together an end-to-end entangled state between Alice and Bob.

\subsection{Constructing the clusters at the major nodes}\label{sec:creatingclique}
The cluster as described above, prepared at each major node in every clock cycle, is pieced together by fusing single photons into progressively larger cluster fragments, probabilistically, using linear-optical circuits and photon detectors. The optimal algorithm for creating photonic cluster states using linear optics---in terms of minimizing the total number of photons consumed and maximizing the eventual probability of success---is not known even for a general $N$-node line cluster. With losses from sources detectors and waveguides during cluster construction, finding the optimal recipe becomes even harder. One design knob is the number of redundant cluster fragments attempted at each step. A higher number of attempts improves the probability of successfully creating the final cluster, but with a higher number of required photon sources and detectors. We refer to this trick of attempting the creation of multiple identical cluster fragments at each step of the process as {\em multiplexing}.

We now describe the resource counts and success-probability calculations for two methods to create the cluster at the major node. The first one is a method implied by previous rough estimates of the resource requirements~\cite{2015.PRX.Li-Benjamin.RescostFTLOQC, 2008.PRL.Varnava-Rudolph.PhotonLoss-LOQCscaling}. We then discuss an improved scheme that decreases the resource requirements during the creation process. Fig.~\ref{multipleximprove} provides a schematic for these two schemes, which we refer to in the discussion below.
 
%Previous studies have calculated or estimated the resource requirements to build clusters based on the average number of attempts required for each probabilistic step \cite{2015.PRX.Li-Benjamin.RescostFTLOQC, 2008.PRL.Varnava-Rudolph.PhotonLoss-LOQCscaling}. In order to create the clusters with high probability on every clock cycle while minimizing the loss associated with photons waiting in waveguides, the attempts at probabilistic operations must be made simultaneously. Furthermore, since the protocol requires successful creation of clusters at every repeater node, the resources available at every repeater station need to significantly greater than the number which would allow for successful cluster creation``on average". We first describe the process of creating the cluster with the probabilistic steps being performed in parallel. We then present improvements to the creation process which decrease the resource requirements of the creation process.

\begin{figure}[h]
    \includegraphics[width=\columnwidth]{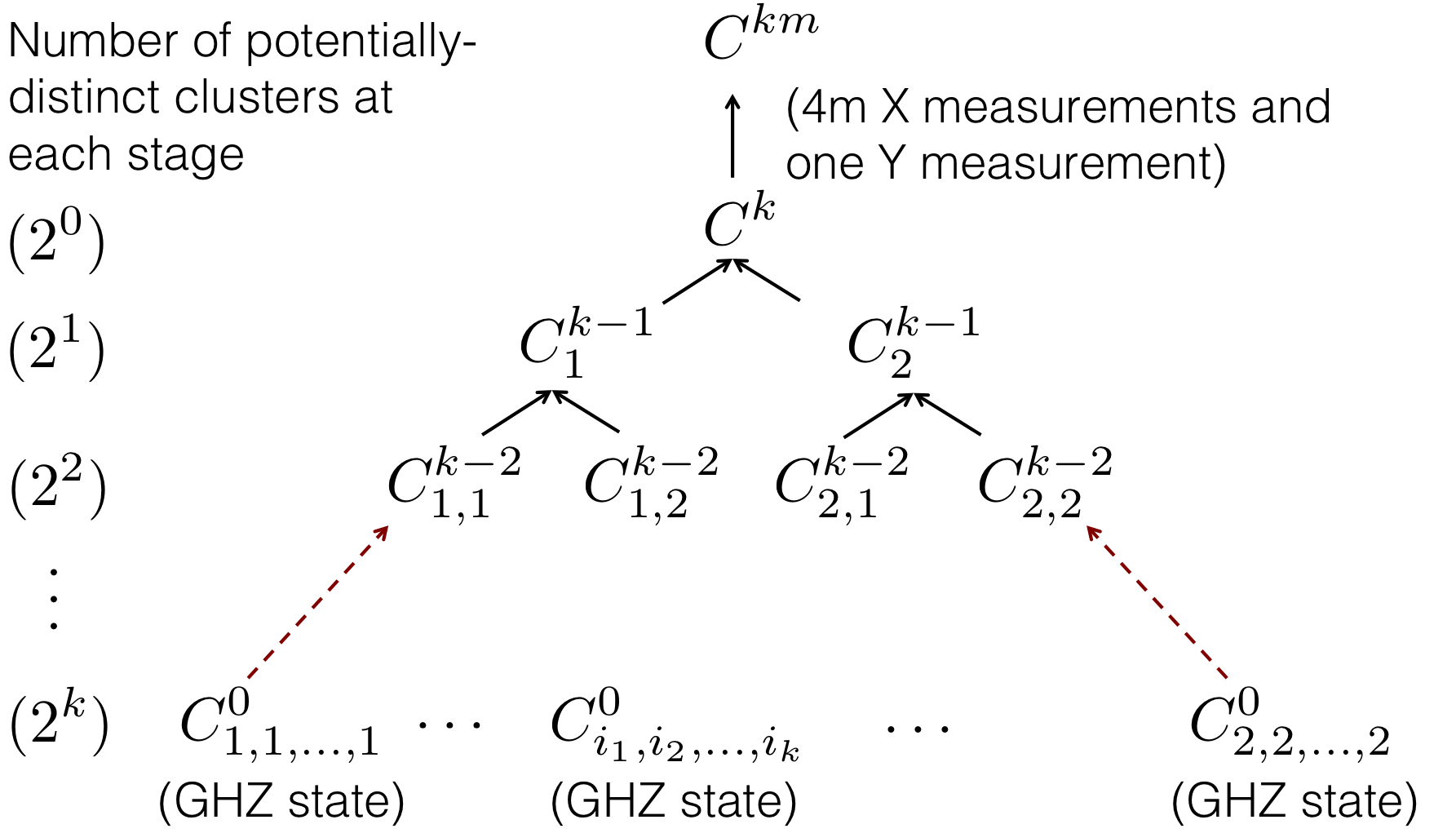}
    \caption{The tree cluster $C^k$ (and the final cluster $C^{km}$ after the $X$ and $Y$ measurements), shown in Fig.~\ref{Lorepeaterconcept}, are created by a sequence of probabilistic linear-optical fusion-II operations, starting from $3$-photon maximally-entangled (GHZ) states.}
    \label{fig:clusters_to_combine}
\end{figure}
Let us label the final cluster $C^{km}$ (see Fig.~\ref{Lorepeaterconcept}) where the letter $m$ signifies that the $Y$ measurement required to turn the inner qubits of the star into a clique (a fully interconnected graph) and the $X$ measurements required to connect the error protection trees to the inner qubits have already been applied. Before these measurements, the (tree) cluster is labelled as $C^{k}$. We label the daughter clusters that are fused together to create $C^k$ as $C_{1}^{k-1}$ and $C_{2}^{k-1}$. The daughter clusters that are fused together to create $C_{1}^{k-1}$ are: $C_{1,1}^{k-2}$ and $C_{1,2}^{k-2}$. The clusters that are fused together to create $C_{1,2}^{k-2}$ are: $C_{1,2,1}^{k-3}$ and $C_{1,2,2}^{k-3}$, and so on (See Fig.~\ref{fig:clusters_to_combine}). At the bottom of the stack are $3$-photon GHZ states, $C_{{\boldsymbol i}}^0$ with ${\boldsymbol i} \equiv i_1,i_2\ldots,i_k$, which are in turn created by groups of $6$ photons fed into linear-optical circuits that generate the $3$-photon GHZ states with probability $P_{\rm GHZ} = \left[\eta_s\eta_d(2-\eta_s\eta_d)\right]^3/32$~\cite{2008.PRL.Varnava-Rudolph.PhotonLoss-LOQCscaling}. The loss rate of the heralded GHZ states is, $1-{\eta}_{\rm GHZ}$ where $\eta_{\rm GHZ} = \eta_s\eta_d/(2-\eta_s\eta_d)$~\cite{2008.PRL.Varnava-Rudolph.PhotonLoss-LOQCscaling}. 

We assume that the cluster $C^k$ can be prepared in a series of $k$ fusion steps, where at each step, clusters of roughly equal sizes are fused together, thus roughly doubling the cluster size in each step~\cite{2015.PRX.Li-Benjamin.RescostFTLOQC}. This assumption becomes accurate in the limit of large clusters. This method ties the final size of the intended cluster ($N_{\rm cluster} = 2^k + 2$ photons) to the number of fusion steps ($k$), and this relationship becomes increasingly exact as $k$ becomes large.  In other words, we assume that $C^{l-1}_{{\boldsymbol i},1}$ and $C^{l-1}_{{\boldsymbol i},2}$ are two clusters each of $p$ photons, which when fused successfully using a fusion-II gate (applied to one photon each of the above two clusters) creates the $2p-2$ photon cluster $C^{l}_{{\boldsymbol i}}$, ${\boldsymbol i} \equiv i_1,i_2\ldots,i_{k-l}$. Starting with the $3$-photon GHZ states $C_{i_1,i_2\ldots,i_k}^0$, the size of $C^k$ is $2^k + 2$ photons. Hence, the minimum number of fusion steps required to build a $N_{\rm cluster}$ photon cluster is $k = \left\lceil \log_2(N_{\rm cluster}-2) \right\rceil$. The label $k$, the number of fusion-II steps used to arrive at $C^k$, also translates to the resource requirements, and the loss rate of each photon in the final cluster, as we show below. Note that $k$ is a function of the branching vector ${\vec b}$ of the error-correction trees used. The larger the error-correction trees, the larger is the final cluster $C^k$, and the larger is the number of steps $k$ required to prepare that cluster.

\subsubsection{The naive multiplexing scheme}

\begin{figure*}
    \includegraphics[width=\textwidth]{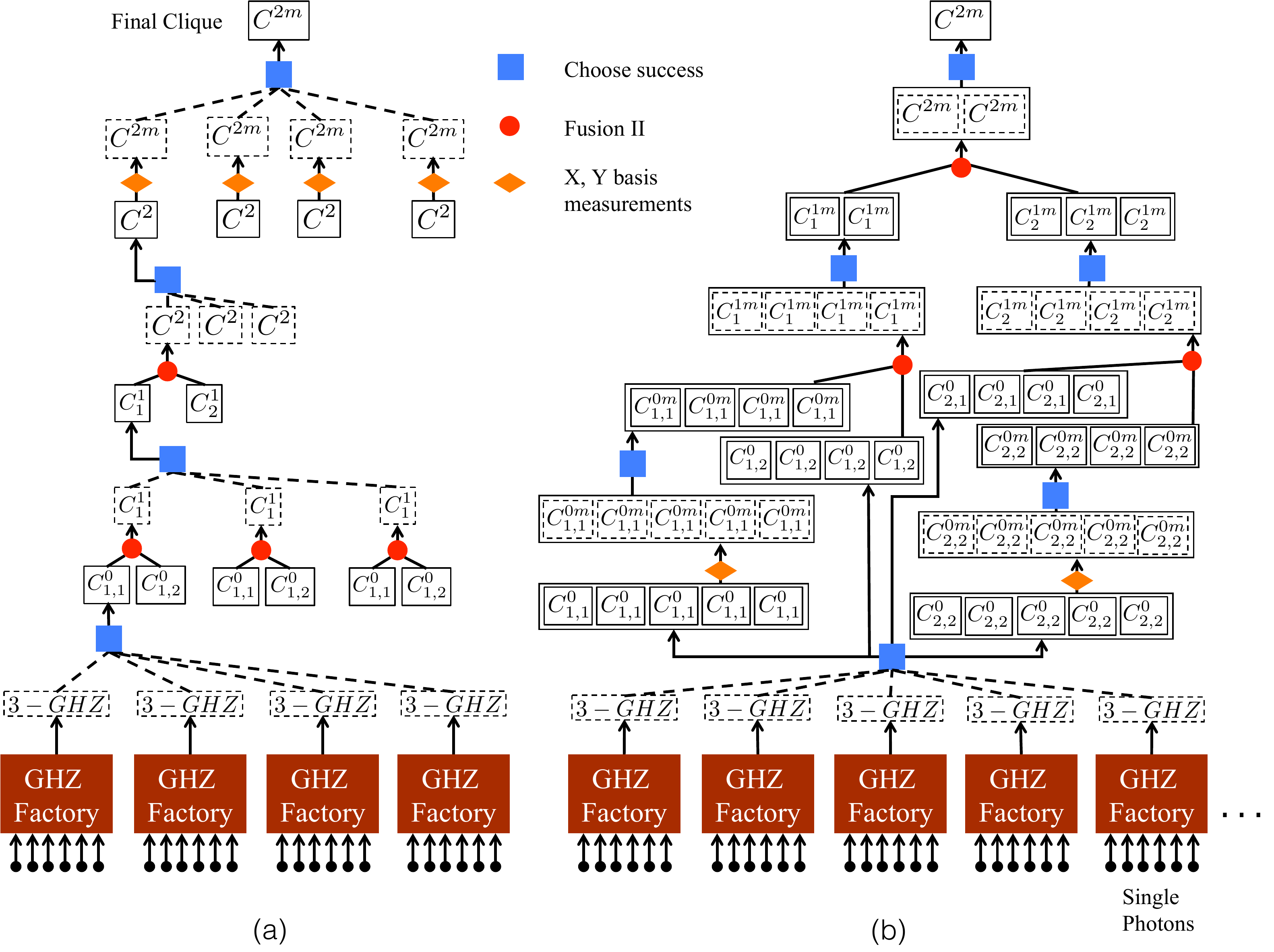}
    \caption{(a) the naive multiplexing scheme. A dashed rectangle represents a cluster that has some probability of having been been created after a probabilistic fusion step (red circle) or at the output of creating GHZ states using linear optics starting from six single photons (labeled `GHZ Factory'). A solid rectangle represents a cluster state that is successfully created with high probability by choosing a successful outcome (blue square) out of several identical copies attempted (dashed boxes). (b) the improved multiplexing scheme. A box surrounding clusters of the same type represents a bank of clusters and any operation applied to the bank is applied to all the clusters in it.}
    \label{multipleximprove}
\end{figure*}

Let us now examine the cluster creation process (depicted for $k=2$ in Fig. \ref{multipleximprove}(a)). At every point we need the cluster fragment $C^l_{{\boldsymbol i}}$, we attempt to create $n_B$ copies of that identical cluster ($n_B=3$ shown in Fig. \ref{multipleximprove}(a)), of which hopefully one is successfully created and heralded for further use. Therefore, creating one usable copy of $C^k$ requires $(2n_B)^k$ GHZ states $C_{i_1,i_2\ldots,i_k}^0$ at the bottom of the stack. Each GHZ state is picked from $n_{\rm GHZ}$ parallel-attempted GHZ states ($n_{\rm GHZ}=4$ shown in Fig. \ref{multipleximprove}(a)), and creating each GHZ state requires $6$ single photons. Therefore, creating one usable copy of $C^k$ requires $(2n_B)^k \times 6n_{\rm GHZ}$ single photons. Finally, at the top of the chain, we create $n_{\rm meas}$ copies of $C^k$ in parallel ($n_{\rm meas}=4$ shown), on each of which the $4m$ $X$ measurements and one $Y$ measurement are performed, to prepare copies of the final required cluster $C^{km}$. We choose $n_{\rm meas}$ such that we obtain with high probability one successfully-created copy of $C^{km}$. Therefore, the total number of single photon sources (shown by black dots at the bottom of Fig.~\ref{multipleximprove}(a)) that need to simultaneously fire on every clock cycle, $N_s = 6n_{\rm GHZ}\,n_{\rm meas}(2n_B)^k$.

The probability of successfully creating a GHZ state $C_{i_1,i_2\ldots,i_k}^0$ is $P_0 = 1 - (1-P_{\rm GHZ})^{n_{\rm GHZ}}$. The success probability of fusion at the $l$-th step---i.e., that of combining $C^{l-1}_{{\boldsymbol i},1}$ and $C^{l-1}_{{\boldsymbol i},2}$ into $C^{l}_{{\boldsymbol i}}$---is given by $Q_l = ({\eta}_{\rm GHZ}P_{\rm chip}^l)^2/2$. The success probability of heralding one cluster $C^{l}_{{\boldsymbol i}}$ (from the $n_B$ parallel copies attempted) is given by the recursive formula, $P_l = 1-(1 - P_{l-1}^2Q_l)^{n_B}$, with $P_0$ given as above. The $4m$ $X$ measurements and one $Y$ measurement required to convert $C^k$ to the final cluster $C^{km}$ succeed with probability $P^\prime = \left({\eta}_{\rm GHZ}P_{\rm chip}^{k+1}\right)^{4m+1}$. Since this step is multiplexed over $n_{\rm meas}$ parallel attempts, the success probability of heralding one copy of the final cluster at a major node is given by, $P_{c1} = 1-(Q_k P^\prime)^{n_{\rm meas}}$. The success probability of all $n$ repeater nodes creating the clusters $C^{km}$ locally during any given clock period, is $P_{cn} = P_{c1}^n$. The blue (dashed) plot in Fig.~\ref{PcnvsNs} shows $P_{cn}$ as a function of $N_s$ for $n=250$ repeater stations (major nodes), $k=7$, and for device parameters as given in Table~\ref{deviceparamtable}.

\subsubsection{The improved multiplexing scheme}

The improved multiplexing scheme we now describe addresses the following deficiencies of the scheme described above.

\begin{itemize}

\item The protocol presented above does not make the most optimal use of the multiple copies of identical clusters that are successfully created at a given step. To illustrate this point, let us consider the $n_B=3$ copies of (attempted) $C^{2}$ clusters that are shown in Fig.~\ref{multipleximprove}(a), of which one successfully created $C^2$ is picked. The first of those three attempted $C^{2}$ clusters is shown to be created by fusing a $C_{1}^1$ cluster and a $C_{2}^1$ cluster. The $C_{1}^1$ is chosen out of $n_B=3$ copies of (attempted) $C_{1}^1$ clusters, as shown. If two of those three copies of $C_{1}^1$ are actually successfully created, the second success goes waste. Note however that the second and the third (of the three attempted) $C^2$ clusters also each need to be created by fusing a $C_{1}^1$ and a $C_{2}^1$. Those two $C_{1}^1$ clusters are also picked from $n_B=3$ copies each of (attempted) $C_{1}^1$ clusters (not shown in the figure). It is thus simple to see that at each time step, a total of $(n_B)^{k} = 9$ copies of $C_{1}^1$ are attempted, but the selection of successes only happen within groups of three, which is clearly inefficient. A far more efficient approach is to maintain one single ``bank" of copies of $C_{1}^1$ and similarly one single bank for copies of $C_{2}^1$, and attempt fusions on clusters from these two banks pairwise (and throw away the excess clusters in the bank that has more copies), to produce a single bank of $C^2$ clusters. This way, one does not have to choose the multiplexing numbers $n_B$, $n_{\rm GHZ}$ and $n_{\rm meas}$, and the total number of single photons $N_s$ directly translates to an overall probability of success $P_{c1}$ of creating the final cluster $C^{km}$. In general, we maintain single banks of each distinct cluster fragment consumed in the entire stack shown in Fig.~\ref{fig:clusters_to_combine}, and for each fusion step shown in Fig.~\ref{fig:clusters_to_combine}, we apply pairwise fusion to {\em all} cluster copies from the two banks corresponding to the two daughter clusters (and throw away the excess clusters from the bank that has more).

\item The $X$ and $Y$ measurements that were performed at the very end (on $4m+1$ nodes of the tree cluster $C^k$, to convert it to the required final cluster $C^{km}$) can be performed at the very beginning---on the appropriate photons (which would eventually become those $4m+1$ photons in $C^k$)---while they are still part of the $3$-photon GHZ states, i.e., before any of the fusion-II operations begin. Making these measurements at the bottom of the stack makes failures much less costly, which in turn significantly reduces the resource requirements (i.e., the $N_s$ required to achieve a given final success probability $P_{\rm cn}$). Appendix \ref{app:reordering} rigorously explains why these measurements can be done on the photons while they are still parts of the GHZ states.

\item The success probability of each of the fusion-II operations (at all $k$ steps in the cluster creation process) can be improved from $1/2$ to $3/4$ by injecting ancilla single photons~\cite{2014.PRL.Ewert-Loock.boostedfusion}. These success probability numbers diminish with source and detection inefficiencies. But, the cost of using additional photons needed (as ancillas) to realize these {\em boosted} fusion gates is far outweighed by the effect of the success-probability improvement, thereby improving the effective tradeoff between $N_s$ and $P_{\rm cn}$.

\end{itemize}

We start with $N_s$ photons and send them all through GHZ factories, hence attempting the creation of $\lfloor N_s/6 \rfloor$ $3$-photon GHZ states. The number of GHZ states $x$ successfully created follows a binomial distribution $B(x,\lfloor N_s/6 \rfloor,P_{\rm GHZ})$ where $B(x,n,p) = {n \choose x} p^x(1-p)^{n-x}$. Hereonafter, let us follow an illustrative set of numbers for a $k=2$ cluster, which is depicted schematically in Fig.~\ref{multipleximprove}(b). Suppose we get $x = 18$ successfully-created GHZ states. These GHZ states are now split into $4$ banks corresponding to $C^0_{1,1}$, $C^0_{1,2}$, $C^0_{2,1}$ and $C^0_{2,2}$. Out of these, let us say $C^0_{1,1}$ and $C^0_{2,2}$ consist of photons that would be eventually measured in $C^k$. As discussed in Appendix \ref{app:reordering}, these qubits can be measured now. Since the measurement of photons has a success probability $P_{\rm chip}{\eta}_{\rm GHZ}$, the number of $C_{1,1}^{0m}$ cluster states ($x$) created as a result of making measurements on $y$ $C_{1,1}^{0}$ states follows a binomial distribution $B(x, y, P_{\rm chip}\eta_{\rm GHZ})$. The banks corresponding to $C^0_{1,1}$ and $C^0_{2,2}$ are given a fraction $1/(P_{\rm chip}{\eta}_{\rm GHZ})$ more GHZ states. Hence, these banks have $5$ GHZ states each whereas the other two have $4$ each. Suppose that measuring the $5$ copies of $C^{0}_{1,1}$ results in $4$ copies of $C^{0m}_{1,1}$, and measuring the $5$ copies of $C^{0}_{2,2}$ results in $4$ copies of $C^{0m}_{2,2}$. The first fusion step is now attempted (i.e., fusing $C^{0m}_{1,1}$ with $C^{0}_{1,2}$, and fusing $C^{0}_{2,1}$ with $C^{0m}_{2,2}$) resulting in $2$ successfully created copies of $C^{1m}_{1}$ and $3$ copies of $C^{1m}_{2}$ (the maximum possible number of successes in both cases was $4$). In the final step, there are $2$ fusion attempts from which we get one copy of the final cluster state $C^{2m}$. 

In general, in a level-$l$ fusion step in Fig.~\ref{fig:clusters_to_combine}, and with $y_1$ and $y_2$ copies in the respective banks of the two daughter clusters, the distribution of the number $x$ of fused states $C^l_{\boldsymbol i}$ is, $B\left(x, \min\{y_1, y_2\}, p_l\right)$, where $p_l = \mu_l^2\left(\frac{1}{2}(\eta_s\eta_d)^2+\frac{1}{4}(\eta_s\eta_d)^4\right)$~\cite{2014.PRL.Ewert-Loock.boostedfusion} and $\mu_l = {\eta}_{\rm GHZ}P_{\rm chip}^{l+1}$ is the survival rate of photons up to before the $l^{th}$ fusion step. The success probabilities of this scheme, $P_{c1}$ (and $P_{cn}$) are calculated using Monte Carlo simulations.

\begin{figure}
    \includegraphics[width=0.9\columnwidth]{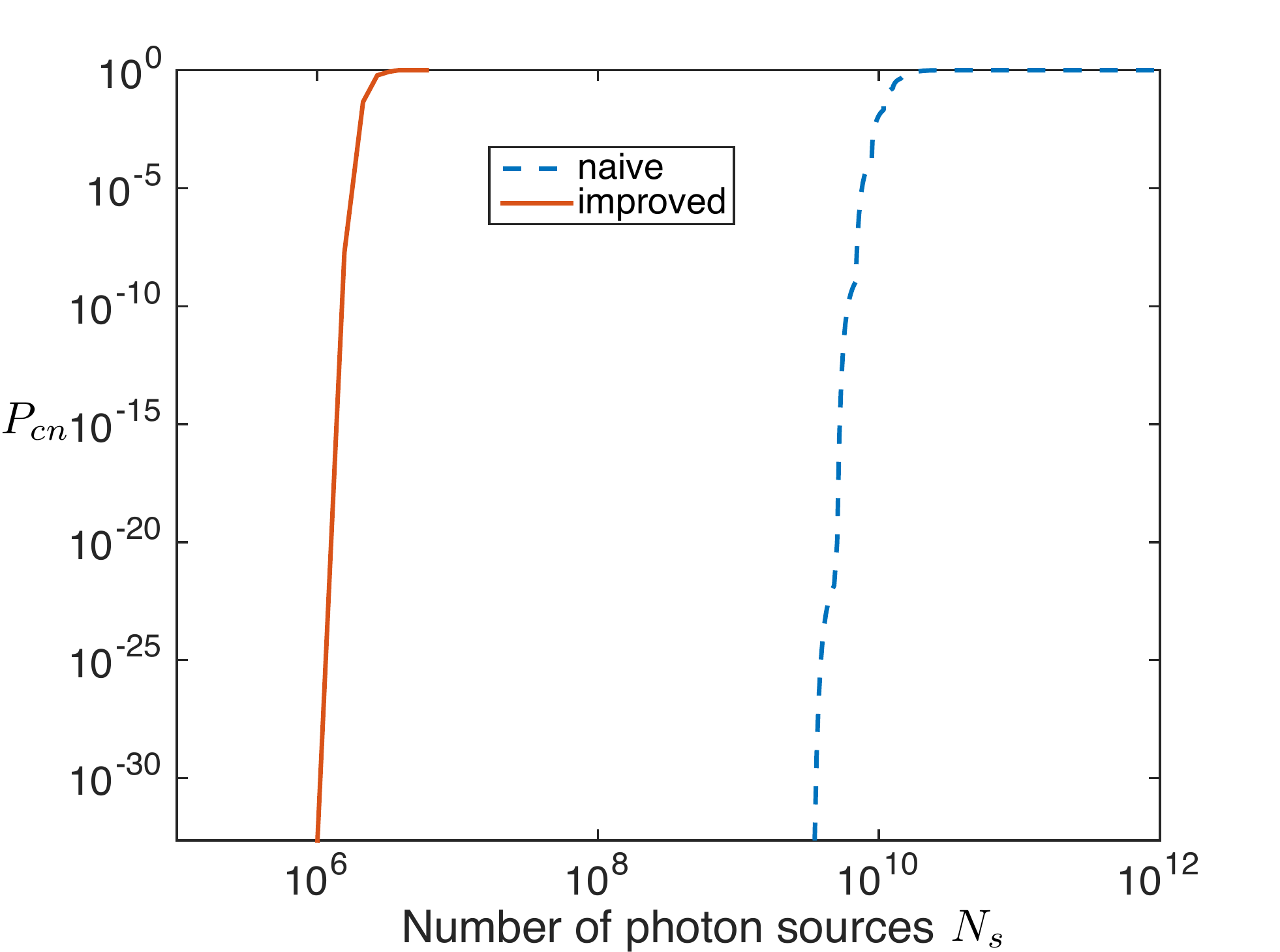}
    \caption{The probability that all $n = 250$ major nodes are simultaneously successful in creating clusters of size $k=7$ fusion steps (i.e., $2^k + 2 = 130$ photon clusters), using the naive and the improved multiplexing schemes.} 
    \label{PcnvsNs}
\end{figure}
In Fig.~\ref{PcnvsNs}, we plot the probability $P_{cn}$ of successfully building clusters $C^k$ (with $k=7$), simultaneously at $n = 250$ major nodes, for both schemes. $n_B$, $n_{\rm GHZ}$ and $n_{\rm meas}$ are optimized for the naive scheme to maximize $P_{cn}$ for any given $N_s$. The plot clearly shows that the improved scheme leads to resource savings by a factor of $\sim10^4$. We further observe that, for both schemes, $P_{cn}$ undergoes a rapid percolation-like transition from zero to one as $N_s$ is increased beyond a certain threshold value. $P_{cn}$ is only a function of $k$, $n$, and $N_s$. We fix $P_{cn} = 0.9$ and calculate the corresponding minimum $N_s$ required, for every value of $k$ and $n$. This sharp-transition behavior of $P_{cn}$ allows us to conveniently split the problem of designing the repeater architecture into two parts: 

(1) choosing an error-protection level by choosing $m$ (number of parallel qubit channels) and $\vec b$ (the branching vector of the error protection trees), which gives us $k$ (indicative of the total cluster size), and using this to calculate the key rate vs. distance achieved---both with $n$ repeater stations, and also the resulting envelope over all $n$; and

(2) given the design choices ($m$ and $\vec b$), calculating the number of photon sources $N_s$ so as to achieve a close-to-unity $P_{cn}$ (probability that all $n$ nodes create the required clusters on every clock cycle), for a given value of $k$ (cluster size at each repeater node), and $n$ (the number of repeater nodes).

\subsection{Measuring the clusters and connecting the chain}\label{sec:measuringclusters}
Once the clusters are created, the outer qubits are sent to minor nodes at the middle of the elementary links, as shown by the arrows in Fig. \ref{Lorepeaterconcept}(c). %The $n$ repeaters are placed a distance $L_0$ apart and an entangled pair separated by $L = nL_0$ is created when the scheme is successful. 
The outer qubits are measured in the Bell basis at the minor nodes using ancilla-assisted boosted fusion gates~\cite{2014.PRL.Ewert-Loock.boostedfusion}. The loss rate seen by the outer qubits is $\epsilon_{\rm trav} \equiv 1-\eta^{\frac{1}{2n}}P_{\rm chip}^{k + 2}{\eta}_{\rm GHZ}\eta_c$ where $\eta^{1/2n}$ is the transmissivity of half of an elementary link (of range $L/2n$). All the physical qubits corresponding to the inner (logical) qubits are stored locally in a fiber bundle with the same attenuation as the communication fiber between the repeater stations. Due to the classical-communication delay, the core qubits see more loss than the outer qubits do, which we define as $\epsilon_{\rm stat} = 1-\eta^{\frac{1}{n}}P_{\rm chip}^{k + 2}P_{\rm fib}{\eta}_{\rm GHZ}\eta_c$. However, it is important to note that, just like in the architecture of~\cite{2014.PRL.Sinclair-Tittel.AtomFreqCombRep, 2015.PRA.Guha-Tittel.QRRateLossAnalysis}, this delay only leads to a latency in the scheme and does not affect the clock rate of the system. 

When the result of the BSMs on the $m$ qubit channels at the two neighboring minor nodes arrive back at a major node, the major node picks one successful qubit channel on either side (if none of the $m$ BSMs were a success on any one of the sides, then that time period is an overall failure). The logical inner qubits corresponding to all the outer qubits that are not deemed part of the successful BSMs are removed from the cluster by measuring them in the $Z$ basis~\cite{2006.PRL.Varnava-Rudolph.CountEC} (note that this $Z$ measurement is a logical one, which benefits from the loss-protection trees). On the two logical qubits (one on either side) corresponding to the successful channels, $X$ basis measurements are performed, which has an effect of extending the entanglement. Alice and Bob, simultaneous with the minor node BSMs, detect the $m$ outer photons sent to them by the first and the last major node in the repeater chain, over links of length $L_0/2$, using one of two randomly-chosen mutually-unbiased bases. Assuming the clusters at all $n$ repeater nodes were successfully created (which happens with probability $P_{cn}$), the conditional probability of generating an end-to-end entangled pair between Alice and Bob, in one clock cycle, is given by the probability that all $n-1$ minor nodes herald at least one successful BSM, and all the pruning logical $X$ and $Z$ measurements on the clusters at all $n$ major nodes are successful, and Alice and Bob both obtain successful detects on at least one of the $m$ qubit channels:
\begin{equation}
P_{\rm meas} = P_Z^{2(m-1)n}P_X^{2n}\left[1-(1-P_B)^m\right]^{n-1}P_{\rm end}^2,
\end{equation}
where $P_X$ and $P_Z$ are the probabilities of successful $X$ and $Z$ basis measurements on the logical inner qubits, respectively. $P_{\rm end}$ is the probability that Alice (resp., Bob) obtains at least one successful detection in one of the $m$ qubit channels.

We quantify the performance of the repeater architecture in terms of the number of shared secret bits generated per mode (i.e., per clock cycle per spatial channel, where $m$ is the number of spatial channels employed). Since, the channel noise comprises of only photon loss, the success probability divided by the number of spatial channels per attempt is the secret key rate (in bits per mode) generated by this scheme, i.e., $R = P_{cn}P_{\rm meas}/2m$ bits/mode. Note that the bits per mode is obtained by dividing by the number of spatial channels that is twice the number of qubit channels ($2m$). This is because we assume single-polarization dual-rail encoding where each qubit on any given spatial channel occupies two successive temporal modes.

\section{Rate calculations}\label{sec:rates}
In this Section, we evaluate the secret key rate achievable using the all-optical repeater architecture described above, while accounting for all the device and channel losses. We first evaluate an expression for $R^{(m, {\vec b})}_n(L)$, the bits-per-mode rate for a given choice of design parameters: $m$ (the number of parallel channels) and $\vec{b}$ (branching vector of the error-protection trees). $L$ is the Alice-to-Bob range and $n$ is the number of equally-spaced repeater nodes that are deployed between Alice and Bob. We evaluate the rate-vs.-distance envelope $R^{(m, {\vec b})}(L)$---the maximum of $R^{(m, {\vec b})}_n(L)$ at any $L$ over the choice of $n \in \left\{1, 2, \ldots\right\}$---and we show explicitly for when $\vec b$ is a depth-$2$ tree, that $R^{(m, {\vec b})}(L) \ge D\eta^s$, with $D$ a constant, $\eta = e^{-\alpha L}$ and $s$ strictly less than $1$. We find by numerical evaluation that this lower bound is tight. We compare this rate-distance envelope with the best rate achievable without the use of quantum repeaters $R_{\rm direct}(L) = -\log_2(1-\eta)$, for some $(m, \vec{b})$ pairs.

A given choice of $m$ and $\vec{b}$ determines $k$, the number of fusion steps required to prepare the final cluster $C^{k}$ prepared by each repeater node at every clock cycle, which in turn quantifies the size ($N_{\rm cluster} = 2^{k}+2$ photons) of $C^{k}$. Next, we choose a value of $k$---a single parameter that quantifies the amount of resources we are willing to dedicate to each repeater node---, and numerically optimize the choice of $m$ and $\vec b$ that is consistent with the chosen $k$, and which maximizes the rate. We denote the rate attainable with $n$ repeater nodes conditioned on the per-node-resource-constraint parameter $k$, as $R^{(k)}_n(L)$ and calculate the optimal rate-vs.-distance envelope $R^{(k)}(L)$ by taking an envelope over the choice of $n$. Finally, we compare the rate-distance envelopes for increasing values of $k$ and translate the values of $k$ to the number of single photon sources required at each repeater node.

The probabilities of fault-tolerant $X$ and $Z$ measurements on one of the (logical) inner qubits of a major node cluster, $P_X$ and $P_Z$, can be expressed in terms of the probabilities $\xi_i$ of a successful `indirect' $Z$ measurement (as described in Section~\ref{sec:prelim}) on a qubit at the $i$-th level of the error-protection tree~\cite{2015.NatureComm.Azuma-Lo.AllOptRep,2006.PRL.Varnava-Rudolph.CountEC}:

\begin{eqnarray}
P_X &=& \xi_0, \, {\text{and}} \\
P_Z &=& (1- \epsilon_{\rm stat} + \epsilon_{\rm stat}\xi_1)^{b_0},
\end{eqnarray} 

\noindent where,

\begin{equation}
\xi_i = 1-\left[ 1-(1-\epsilon_{\rm stat})(1-\epsilon_{\rm stat}+\epsilon_{\rm stat}\xi_{i+2})^{b_{i+1}}\right]^{b_i},
\label{Rkrecur}
\end{equation}
and $i\leq l$, $\xi_{l+1} = 0$, $b_{l+1} = 0$.

Let us assume a tree depth of $d = 2$, i.e., $\vec{b} = \left[b_0 \ b_1\right]$, which is consistent with our numerical findings on the optimal branching vector as described later in the paper (see table~\ref{performancecomparisontable}). For a depth-$2$ branching vector, using Eq.~\eqref{Rkrecur}, we find that $\xi_0 = 1-\left[1-\left(1-\epsilon_{\rm stat})^{b_1+1}\right)\right]^{b_0}$ and $\xi_1 = 1-\epsilon_{\rm stat}^{b_1}$. Thus, 

\begin{eqnarray}
P_X &=& 1-\left[1- \left(\eta^{\frac{1}{n}}\right)^{b_1+1}B^{b_1+1}\right]^{b_0}, \,{\text{and}} \label{eq:PX}\\
P_Z &=& \left[ 1-\left(1-\eta^{\frac{1}{n}}B\right)^{b_1+1}\right]^{b_0},\label{eq:PZ}
\end{eqnarray} 
and the Bell measurement success probability becomes 

\begin{equation}
P_B = \frac{AB^2}{m}\eta^{\frac{1}{n}},
\label{eq:PB}
\end{equation}
where $A = m\left(\frac{1}{2}(\eta_s\eta_d\right)^2+\frac{1}{4}(\eta_s\eta_d)^4)/P_{\rm fib}^2$, $B  = P_{\rm chip}^{k+2}P_{\rm fib}{\eta}_{\rm GHZ}\eta_c$.

The probability of at least one successful detection at Alice's (or Bob's) end is given by

\begin{equation}
P_{\rm end} = 1 - \left(1-\eta^{\frac{1}{2n}}C\right)^m,
\label{eq:Pend}
\end{equation} 
where $C  = P_{\rm chip}^{k+2}{\eta}_{\rm GHZ}\eta_c$.

\begin{figure}[h]
    \includegraphics[width=0.5\textwidth]{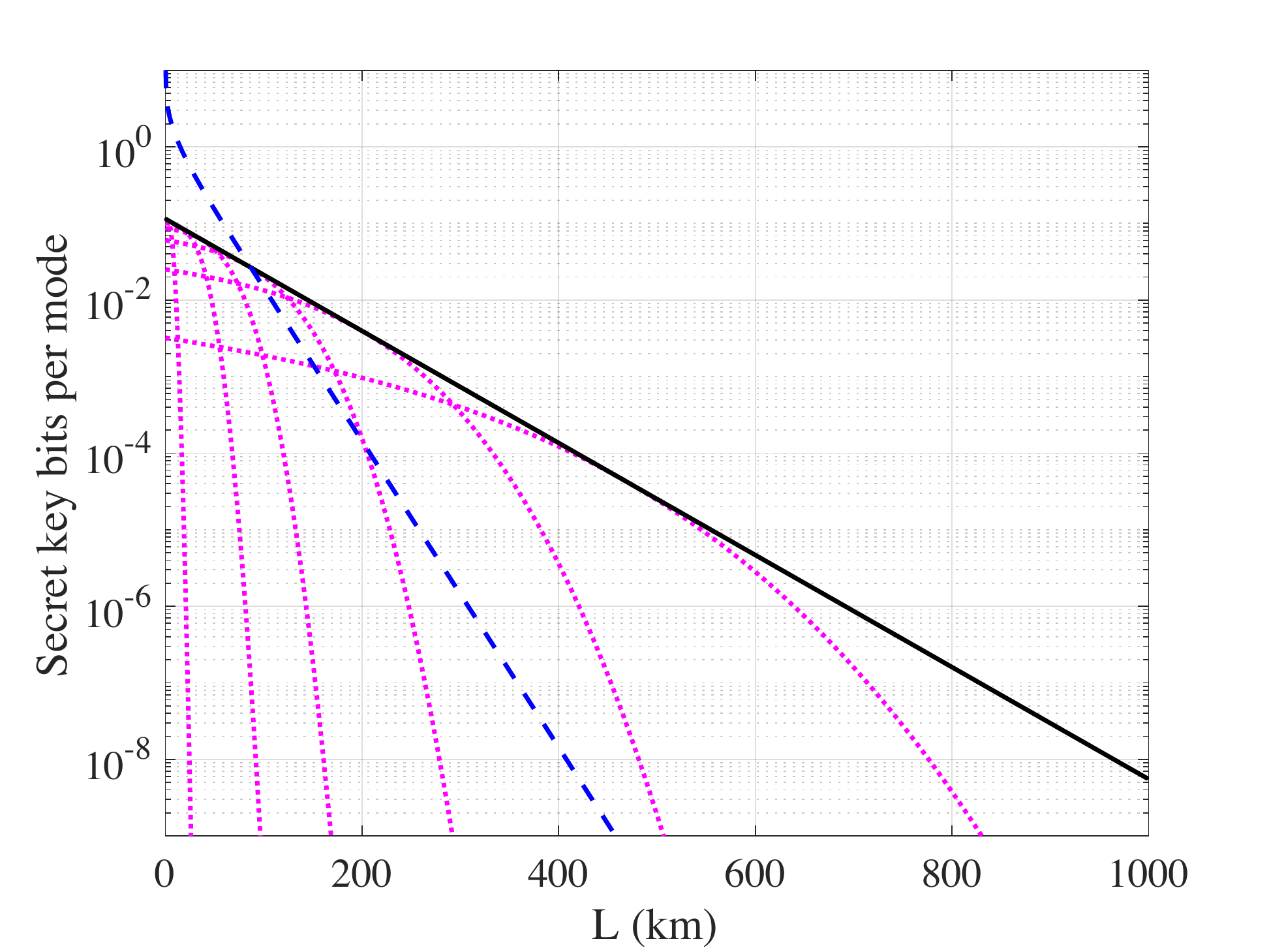}
    \caption{The key rate (in bits per mode) $R_n^{(m,{\vec b})}(L)$ achieved by an $n$-node repeater chain shown as a function of range $L$, for $n = 1, 10, 24, 56, 133$, and $314$ (magenta dotted plots), with $m=4$ parallel channels and $\vec{b} = \left\{7, 3\right\}$ trees. The analytical lower bound to the rate-distance envelope $R_{\rm LB}^{(m,{\vec b})}(L)$ (black solid plot) is seen to surpass the best-possible repeaterless-QKD rate $R_{\rm direct}(L)$ (blue dashed plot) at $L=87$ km.} 
    \label{anscalingplot}
\end{figure}

We now have the bits-per-mode rate achievable with an $n$-repeater-node chain,

\begin{equation}
R_n^{(m, {\vec b})}(L) = \frac{P_{cn}}{2m}P_{\rm end}^2P_Z^{2(m-1)n}P_X^{2n}\left[1-(1-P_B)^m\right]^{n-1},
\end{equation}
with $P_X$, $P_Z$, $P_B$ and $P_{\rm end}$ as given in Eqs.~\eqref{eq:PX},~\eqref{eq:PZ},~\eqref{eq:PB} and~\eqref{eq:Pend}, with $\eta = e^{-\alpha L}$ the transmissivity of the end-of-end channel (of range $L$). See the dotted magenta curves in Fig.~\ref{anscalingplot} for the plots of $R_n^{(m, {\vec b})}(L)$ as a function of $L$ for a few chosen values of $n$. 

One way to obtain a lower bound of the envelope over the plots $R_n^{(m, {\vec b})}(L)$ over all choices of $n$ (see black plot in Fig.~\ref{anscalingplot}), is to pick one point $(L_n, R_n^{(m, {\vec b})}(L_n))$ on each of the rate-distance functions $R_n^{(m, {\vec b})}(L)$, $n=0, 1, 2, \ldots$, and connect them. Let us choose $L_n$ as: 
\begin{equation}
L_n = nz \,{\ln(AB^2)}/{\alpha},
\label{Lnlocus}
\end{equation}
with $z$ being a constant that is yet to be chosen. The Alice-to-Bob channel transmissivity at these range values are therefore given by:
\begin{equation}
\eta_n = e^{-\alpha L_n} = e^{-nz \ln(AB^2)}.
\label{etanzlocus}
\end{equation}
We now evaluate a locus of the (range, rate) pairs $(L_n, R_n^{(m, {\vec b})}(L_n))$ over $n \in \left\{0, 1, 2, \ldots\right\}$ and choose the parameter $z$ we left undetermined in Eq.~\eqref{etanzlocus} so as to maximize the rate-distance envelope. We call this rate-distance envelope $R_{\rm LB}^{(m, {\vec b})}(L)$ since this is by construction a lower bound on the true envelope $R^{(m, {\vec b})}(L)$. 

Let us evaluate $P_X$, $P_Z$, $P_B$ and $P_{\rm end}$ at $\eta = \eta_n$ (i.e., substitute $\eta^{1/n} = \left(AB^2\right)^{-z}$ in the respective expressions) and define the following quantities:
\begin{eqnarray}
p_X &=& 1-\left[1- \left(AB^2\right)^{-z(b_1+1)}B^{b_1+1}\right]^{b_0}, \,{\text{and}} \label{eq:px}\\
p_Z &=& \left[ 1-\left(1-\left(AB^2\right)^{-z}B\right)^{b_1+1}\right]^{b_0},\label{eq:pz}\\
p_B &=& \frac{1}{m}\left(AB^2\right)^{1-z},\label{eq:pb}\,{\text{and}}\\
p_{\rm end} &=& 1 - \left(1-\left(AB^2\right)^{-z/2}C\right)^m,\label{eq:pend}
\end{eqnarray} 
using which let us define the following: $q_1 = p_Z^{2(m-1)}p_X^2$, $q_2 = 1-(1-p_B)^m$, and $q_3 = p_{\rm end}^2$, and obtain:
\begin{equation}
R_n^{(m, {\vec b})}(L_n) = (q_1 q_2)^{n}\frac{q_3P_{cn}}{2mq_2}.
\label{eq:R_Ln}
\end{equation}
To obtain the envelope $R_{\rm LB}^{(m, {\vec b})}(L)$, we need to calculate the locus of the distance-rate pairs $(L_n, R_n^{(m, {\vec b})}(L_n))$ over $n \in \left\{1, 2, \ldots\right\}$. We do this by eliminating $n$ from Eqs.~\eqref{Lnlocus} and \eqref{eq:R_Ln}. With a little algebra, and expressing the envelope in terms of $\eta = e^{-\alpha L}$, we get the following:
\begin{equation}
R_{\rm LB}^{(m, {\vec b})}(\eta) = D \eta^{s},
\label{eq:LBenvelope}
\end{equation}
where $D = \frac{q_3P_{cn}}{2mq_2}$ and the exponent $s  = -\frac{\ln(q_1q_2)}{z\ln(AB^2)}$. 

Note that $R_{\rm LB}^{(m, {\vec b})}(L)$ in~\eqref{eq:LBenvelope} is a lower bound on the actual rate-distance function $R^{(m, {\vec b})}(L)$ for any value of the parameter $z$ that we left undetermined in our choice of the range values $L_n$ we used to evaluate $R_{\rm LB}^{(m, {\vec b})}(L)$. We numerically optimize the choice of $z$ such that the value of the exponent $s$ is minimized (note that $q_1$, $q_2$ and $q_3$ are all functions of $z$). 

In Fig.~\ref{anscalingplot}, we plot $R_n^{(m, {\vec b})}(L)$ (bits per mode) as a function of $L$ (km) for $n = 1, 10, 24, 56, 133$, and $314$ (magenta dotted plots), with $\vec{b} = \left\{7, 3\right\}$ and $m=4$, and other device parameters as summarized in Table~\ref{deviceparamtable}. These values of $m$ and $\vec b$ translate to $k=8$, i.e., $2^8 + 2 = 258$ photon clusters created at each node at every clock cycle. We also plot the analytical rate-envelope lower bound in Eq.~\eqref{eq:LBenvelope}, $R^{(m,{\vec b})}_{\rm LB}(L)$ (black solid line), with the optimal $z$ computed numerically. For the chosen parameters, we get $D = 0.11$ and $s = 0.37$. The analytical lower bound $R_{\rm LB}^{(m,{\vec b})}(L)$ is visually indistinguishable at the scale of the plot from the numerically-obtained rate-distance envelope $R^{(m,{\vec b})}(L)$. This excellent agreement persists for all values of $m$ and $\vec b$ we have have tried. 

One interesting implication of the range values $L_n$ in Eq.~\eqref{Lnlocus} lying on the rate-distance envelope is that the distance between each repeater (major) node, 
\begin{equation}
L_0 \equiv \frac{L}{n} = \frac{\ln(AB^2)}{\alpha}
\end{equation} 
is a constant and independent of the total range $L$. In other words, given the device parameters and the choice of the major-node cluster size (i.e., $m$ and $\vec b$), there is an optimal gap with which repeaters should be placed---no more, and no less. For the numbers used for the plots in Fig.~\ref{anscalingplot}, $L_0 = 1.49$ km. Fig.~\ref{anscalingplot} also shows $R_{\rm direct}(L)$ for comparison (blue dashed plot), which the repeater scheme is seen to outperform beyond a range of $87$ km.

As shown by the above example, our repeater scheme, even when built with lossy components, can achieve $s < 1$ i.e. it outperforms the optimum repeater-less rate $R_{\rm direct}(L)$. The value of the exponent $s$ achievable by the repeater scheme can be improved (lowered) by enhancing the level of error correction (i.e., choosing a larger $\vec b$). Doing so increases the size of the clusters ($2^k + 2$ photons) needed at each repeater nodes, and hence increases the number of photon sources $N_s$ required locally at each node. In Fig.~\ref{bpmvsdist}, we plot the $R^{(k)}(L)$, numerically-evaluated envelopes of the rate-distance functions $R^{(k)}_n(L)$, parametrized by the single parameter $k$ that quantifies the size of the clusters prepared by the repeaters at each clock cycle. It is seen that the rate-distance exponent $s$ improves (decreases) as $k$ increases.

\begin{table}[h]
\begin{tabular}{ | c | c | p{20mm} |}
    \hline
    Device parameter & symbol & value\\ \hline \hline
    fiber loss coefficient & $\alpha$ & $0.046~\textrm{km}^{-1}$ ($0.2$ dB/km) \\ \hline
    on-chip loss coefficient & $\beta$ & $0.62~\textrm{m}^{-1}$ ($2.7$ dB/m) \\ \hline
    feed-forward time in fiber & $\tau_f$ & $102.85$ ns \\ \hline
    feed-forward time on-chip & $\tau_s$ & $20$ ps \\ \hline
    chip to fiber coupling efficiency & $\eta_{c}$ & $0.99$ \\ \hline
    source detector efficiency product & $\eta_s\eta_d$ & 0.99 \\ \hline
    speed of light in fiber & $c_f$  & $2 \times 10^8 m/s$ \\ \hline
    speed of light on chip & $c_{ch}$ & $7.6 \times 10^7 m/s$ \\ \hline
  \end{tabular}
\caption[Table caption text]{Assumed values for device performance parameters. The source detector efficiency product $\eta_s\eta_d$ is sufficient for the purposes of the calculations in this paper, and need not be specified separately. Recall that $P_{\rm chip} = e^{-\beta \tau_s c_{\rm ch}}$, $P_{\rm fib} = e^{-\alpha \tau_f c_{\rm f}}$, and ${\eta}_{\rm GHZ} = \eta_s\eta_d/(2-\eta_s\eta_d)$. $\tau_f$ has been chosen to make $P_{\rm chip} = P_{\rm fib}$.} 
\label{deviceparamtable}
\end{table}

\section{Discussion}\label{sec:discussion}
In this Section, we go back to the all-photonic repeater architecture proposed by Azuma {\em et al.}~\cite{2015.NatureComm.Azuma-Lo.AllOptRep}, and discuss the main modifications (improvements) we considered in the architecture we described and analyzed above. We also show a comparative study of the resource requirements and rate performance of the naive scheme and our modified scheme. Following are the salient differences between the architecture we analyzed above, and the one proposed in~\cite{2015.NatureComm.Azuma-Lo.AllOptRep}.

{\em Retaining vs. transmitting the clusters}---In the proposal of~\cite{2015.NatureComm.Azuma-Lo.AllOptRep}, all the logical inner qubits, along with the outer qubits (i.e., all the $N$ photons of the cluster at a major node) are sent to the minor node, whereas we store the inner qubit photons in a fiber spool locally at the major nodes. The former has an advantage that no classical communication needs to happen from minor nodes back to major nodes before the logical $X$ and logical $Z$ measurements are done to the logical inner qubits, since all those qubits are present locally at the minor nodes when the BSMs are performed there on outer-qubit pairs from neighboring major node clusters. The advantage of our (latter) scheme is that the number of parallel physical channels needed ($2m$) is much smaller as compared to the number needed ($N$) for the scheme in~\cite{2015.NatureComm.Azuma-Lo.AllOptRep}. For the numbers in Fig.~\ref{anscalingplot}, that is $8$ as opposed to $208$ parallel fiber channels connecting successive repeater nodes. 

\begin{figure}[h]
    \includegraphics[width=0.5\textwidth]{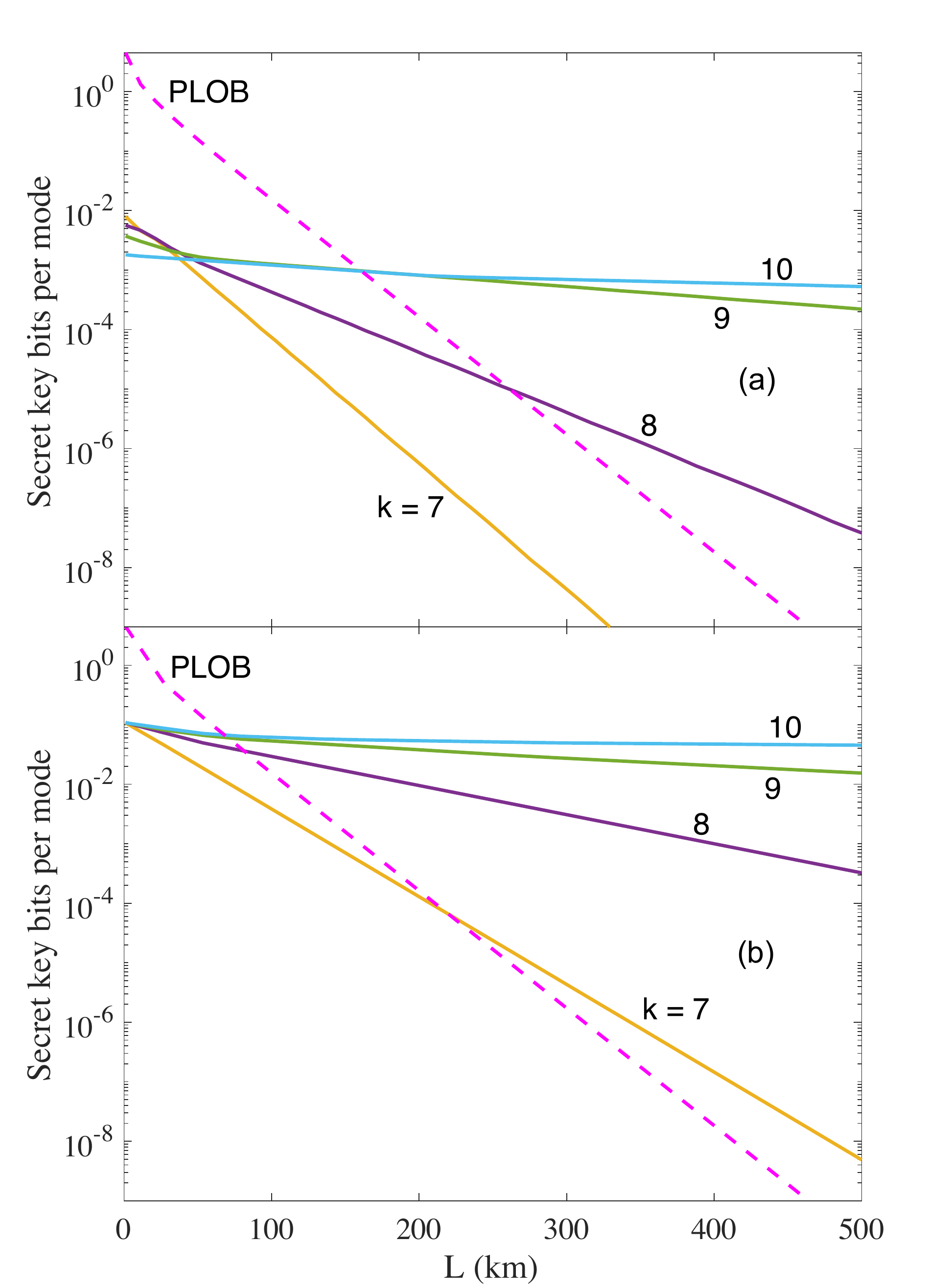}
    \caption{The bits per mode rates $R^{(k)}(L)$ plotted for different values of $k$, the numbers of fusion steps, for the (a) naive scheme and (b) with the improvements of this paper. The repeater-less rate bound $R_{\rm direct}(L)$ is the pink dashed line. $N_{\rm cluster} = 2^k + 2$ is the total number of photons in the cluster generated at each repeater in every clock cycle.} 
    \label{bpmvsdist}
\end{figure}

{\em Difference in the bits-per-mode rate}---Further, the bits per mode achieved by the architecture in~\cite{2015.NatureComm.Azuma-Lo.AllOptRep} would be given by $P_{cn}P_{\rm meas}/N$, whereas the bits per mode achieved by our modified architecture would be $P_{cn}P_{\rm meas}/2m$. The $P_{\rm meas}$ of the former is higher (due to lower loss incurred by the photons of the logical inner qubits of the clusters as they do not need to wait in a lossy fiber spool while waiting for the classical information to fly back from the minor nodes). However, the other improvements described below more than compensate for the better $P_{\rm meas}$, and the latter scheme achieves a far better bits-per-mode performance (see Fig.~\ref{bpmvsdist}).

{\em Linear optic vs. boosted linear optic fusion gates}---We propose the use of the improved Bell-state measurement scheme of Ewert {\em et al.}~\cite{2014.PRL.Ewert-Loock.boostedfusion} that inject four single photons to boost the success probability of the fusion-II gate. Our calculations show that the cost of using these additional ancilla photons is far outweighed by the effect of the improved success probability, in the performance of the repeater architecture, despite assuming lossy sources and detectors.

{\em Improved multiplexing scheme for cluster generation}---We use an improved multiplexing scheme to create the clusters at the major nodes, as described in Section~\ref{sec:creatingclique} and depicted in Fig.~\ref{multipleximprove}(b). Previous studies have estimated the resource requirements for cluster generation based on the average number of attempts required for each probabilistic steps~\cite{2008.PRL.Varnava-Rudolph.PhotonLoss-LOQCscaling, 2015.PRX.Li-Benjamin.RescostFTLOQC}. However, in order to generate the required cluster at every repeater station on every clock cycle with high probability, the resources required at each repeater station need to be greater than the number that would allow for cluster creation ``on average''. To our knowledge, this is the first study that explicitly looks at how probabilistic operations need to be multiplexed in a real system.

{\em Pushing the measurements ahead during cluster creation}---The single qubits measurements that do not depend on the outcomes of Bell measurements at the minor nodes, are performed before the fusion operations, directly on the photons of the GHZ states, very early during the cluster creation process.

Let us now see what the above modifications to the architecture does to the rate performance. The bits-per-mode rates for the naive and the improved schemes are plotted in Fig.~\ref{bpmvsdist}(a) and (b), respectively. We assume device loss parameters as listed in Table~\ref{deviceparamtable} for both sets of plots. In each plot, we compute the rate-distance performance (envelopes taken over $n$, the number of repeater nodes) for  four different error-protection levels (i.e., $k = 7, 8, 9$, and $10$). For every point on each rate-distance envelope, $m$ and $\vec{b}$ are optimally chosen (consistent with the given $k$). Each rate-distance plot exhibits the $D\eta^s = De^{-s\alpha L}$ behavior, and the exponent $s$ diminishes as a higher $k$ is chosen. For the naive scheme, the minimum $k$ for which the repeater can beat $R_{\rm direct}(L)$ (pink-dashed line) is $k = 8$ and the optimized clusters at the major nodes have $192$ photons each. Hence, the scheme would require $208$ parallel fiber links connecting successive nodes. In comparison, in the improved scheme, $k = 7$ is sufficient to beat $R_{\rm direct}(L)$, and requires $2m = 8$ parallel fiber links. The optimal tree depth, for this $k=7$ rate plot is found to be $d = 2$, which is consistent with the analytical development in Section~\ref{sec:rates}.

Table~\ref{performancecomparisontable} lists, at a range of $L=300$ km, and for each of the cases ($k =7, 8, 9, 10$), the optimal values of $m$ for the naive ($m_{\rm naive}$) and new schemes ($m_{\rm new}$), the optimal branching vector for the naive ($\vec{b}_{\rm naive}$) and new schemes ($\vec{b}_{\rm new}$), and the number of parallel fiber links needed in the naive scheme ($N_{\rm naive}$). In the case of the new scheme, the number of parallel fiber links needed is simply $2m_{\rm new}$.

\begin{table}[h]
\begin{tabular}{ | c || c | c | c || c | c |}
    \hline
    $k$ & $m_{\rm naive}$ & $N_{\rm naive}$ & $\vec{b}_{\rm naive}$ & $m_{\rm new}$ & $\vec{b}_{\rm new}$\\ \hline \hline
    $7$ & $5$ & $100$ & $\left\{3, 2\right\}$ &$4$ &$\left\{4, 2\right\}$\\ \hline
    $8$ & $8$ & $208$ & $\left\{4, 2\right\}$ &$5$ &$\left\{5, 3\right\}$\\ \hline
    $9$ & $11$ & $462$ & $\left\{5, 3\right\}$ &$6$ &$\left\{7, 4\right\}$\\ \hline
    $10$ & $12$ & $864$ & $\left\{7, 4\right\}$ &$8$ &$\left\{10, 5\right\}$\\ \hline
  \end{tabular}
\caption[Table caption text]{For $k =7, 8, 9$, and $10$, at $L=300$ km range, $m_{\rm naive}$ and $m_{\rm new}$ are the optimal values of $m$ for the naive and new schemes respectively. $\vec{b}_{\rm naive}$ and $\vec{b}_{\rm new}$ are the optimal values of $\vec{b}$ for the naive and new schemes respectively. $N_{\rm naive}$ is the corresponding number of parallel fiber links needed between successive repeater nodes in the naive scheme. For the new scheme, the number of parallel links is $2m_{\rm new}$.}
\label{performancecomparisontable}
\end{table}

Let us now compare the resources (number of photons, $N_s$) required to build the major node clusters, for the respective cases that can (barely) beat $R_{\rm direct}(L)$. The naive scheme requires $1.9 \times 10^{11}$ photon sources at each major node, while the new scheme requires $3.3 \times 10^6$ sources, an improvement of $5$ orders of magnitude (see Fig.~\ref{PcnvsNs}). It is also interesting to note that if the primitive resources were 3-photon GHZ sources rather than single photon sources, 15 thousand GHZ sources would be required, a relatively smaller number. 

Given the size of the earth, for terrestrial long distance communications, it is useful to quantify the performance of our (improved) all-optical repeater scheme at say $5000$ km. Without quantum repeaters, the best QKD protocol realized with ideal devices cannot exceed a key rate of $2.9 \times 10^{-99}$ bits per mode at this distance. Our all-optical repeater scheme, with $954$-photon clusters ($k = 10$) at each repeater node can attain a key rate of $8 \times 10^{-3}$ bits per mode using $2m = 18$ parallel channels and $n = 12411$ repeater nodes, which translates to a $144$ kHz key generation rate assuming a $1$ MHz repetition rate. If we employed $518$-photon clusters ($k = 9$) instead, the rate achieved would only be $4 \times 10^{-8}$ bits per mode using $2m = 14$ parallel channels and $n = 12255$ repeater nodes. The number of photon sources required at a repeater node to create the required clusters (using linear optics) for the above two example cluster-size constraints are $1.2 \times 10^{8}$ and $3.6 \times 10^{7}$, respectively.

In the presence of losses in the waveguide, there is a maximum sustainable size of the clusters at the major nodes, at least for the error protection methods described in this paper. A larger cluster requires a greater creation time and hence, each photon in the cluster sees a larger effective loss rate (stemming from the $P_{\rm chip}^{k}$ term in $\epsilon_{\rm trav}$ and $\epsilon_{\rm stat}$). Since the error correction scheme has a maximum loss tolerance of $50\%$, there is a maximum size of the clusters that can be created and thus a maximum level of error protection that a qubit can have. So, given a set of device losses, increasing the error protection level (viz., $k$) cannot indefinitely improve the rate performance.

The aforesaid detrimental effect of loss with an increasing cluster size has more serious implications for cluster-state linear optical quantum computing (LOQC) in general, using the tree-based counterfactual error correction technique~\cite{2006.PRL.Varnava-Rudolph.CountEC}. This is because a polynomial scaling of the number of photon sources (with the size of the cluster) is required in the asymptotic limit for the LOQC scheme to be scalable. The failure probability of every qubit needs to decrease exponentially with the size of the computation. Hence, the level of protection of each qubit must increase with the size of the problem, which implies a greater cluster creation time and hence a greater loss rate. Since there is a $50 \%$ ceiling on the tolerable photon loss with the tree code, it is not possible to achieve the required level of protection for arbitrarily large computations, as discussed above for the case of an all-photon quantum repeater. Developing a scalable method for creation of arbitrarily large clusters in constant time would solve this problem and will also allow for a polynomial scaling of the number of photons with computation size. A recent paper proposes using counterfactual error correction to fault-tolerantly create surface code data qubits~\cite{2015.PRX.Li-Benjamin.RescostFTLOQC}. However, the resource requirements for this scheme are extremely high.

\section{Conclusions}\label{sec:conclusions}
In conclusion, we have performed a rigorous analysis of the resource requirements, and the achievable secret key rates of an all-optical repeater scheme that improves upon a recent proposal~\cite{2015.NatureComm.Azuma-Lo.AllOptRep}, while taking into account all the losses in the system. While the all-optical repeater proposal of~\cite{2015.NatureComm.Azuma-Lo.AllOptRep} presents an important conceptual advancement, we show that it may not be practically feasible given its astronomical resource requirements, both in terms of the number of photon sources and detectors needed at each repeater node, as well as the number of parallel optical fiber channels that must connect successive repeater nodes. Our scheme improves the practicality immensely in both of the aforementioned metrics, as well as the actual rate-vs.-distance performance achieved. In particular, the number of photon sources required at each node is reduced by $5$ orders of magnitude, and the number of parallel channels between repeater nodes required to beat the performance of a direct-transmission QKD scheme is brought down from more than two hundred, to $8$. These results suggest that further theoretical improvements on quantum photonic fault tolerant schemes may further improve the performance of all-optical quantum repeaters, as well as other applications of all-optical quantum processing. One of our major contributions in this paper was to rigorously prove that the rate-loss scaling by the aforementioned genre of all-optical quantum repeaters with a fixed cluster size is given by $R = D\eta^s$ bits per mode, where $D$ and $s$ are constants that are functions of various device loss parameters, and that of design choices made (to choose the level of error protection). The fact that it is possible to achieve a value as the exponent $s < 1$ proves the fact that this scheme can outperform the key rates attainable by any QKD protocol that does not employ quantum repeaters, the rate performance of which are upper bounded by $R_{\rm direct}(\eta) \approx 1.44\,\eta$ for $\eta \ll 1$, whose linear rate-transmittance decay implies $s=1$.

In future work, it will be interesting to incorporate more realistic effects into the resource-performance tradeoff calculations of all-optical repeaters, in particular mode-mismatch errors in the passive interferometric manipulations on the photons held locally at the repeaters, and multi-photon errors arising from imperfect sources and noisy detectors. Finally, it would be instructive to analyze and compare other forms of quantum repeater architectures, especially forward-error-corrected one-way transmission schemes~\cite{2015.ArXiv.Muralidharan-Jiang.RepeaterGen}, realized only with flying photons, linear optics and detectors, but no quantum memories.

\begin{acknowledgments}
This research was funded by the DARPA project {\em Scalable Engineering of Quantum Optical Information Processing Architectures} (SEQUOIA), under US Army contract number W31P4Q-15-C-0045 and by the Air Force Office of Scientific Research MURI (FA9550-14-1-0052). We would like to thank Sreraman Muralidharan, Liang Jiang, Darius Bunandar, Koji Azuma, Hoi-Kwong Lo and Stefano Pirandola for helpful discussions.
\end{acknowledgments}

\appendix

\section{Re-ordering measurements in the cluster-creation process}\label{app:reordering}
\begin{figure}[h]
    \includegraphics[width=0.5\textwidth]{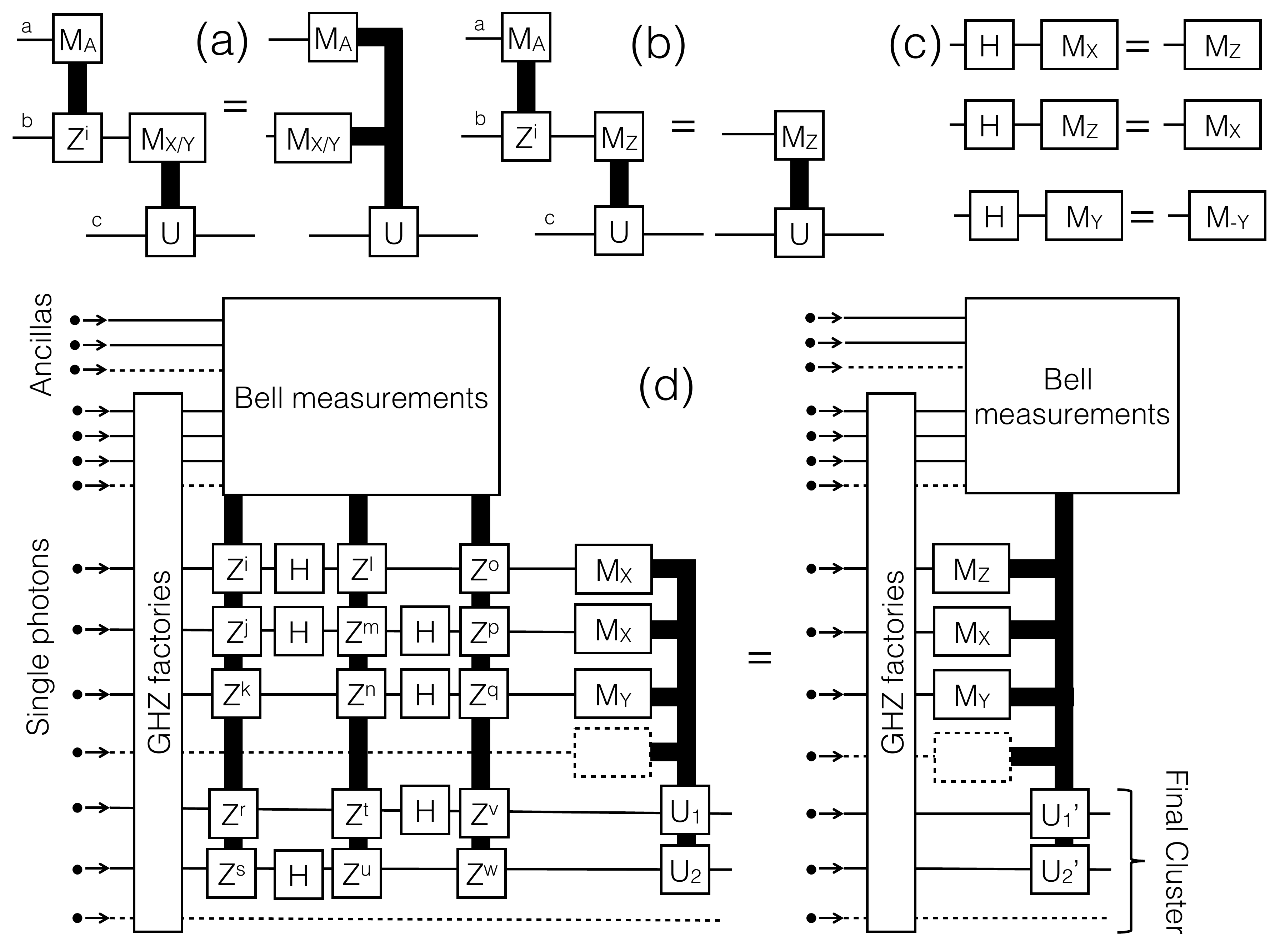}
    \caption{Single qubit measurements can be applied before fusion operations. (a) $X$ and $Y$ basis measurements can be moved before conditional $Z$ operators. (b) $Z$ operators before $Z$ basis measurements can be removed. (c) Hadamard gates followed by measurement in the $X$, $Y$ or $Z$ basis is equivalent to direct measurement in a different pauli basis. (d) Single qubit measurements on the final cluster can be moved before fusion operations.} 
    \label{fig: measureearly}
\end{figure}

In this Section, we explain why the $X$ measurements required to attach trees for counterfactual error correction and the $Y$ measurement required to create the ``clique" from the ``star" cluster can be applied before the fusion operations. This makes the cluster creation process more efficient. The reordering of the operations is depicted in Fig.~\ref{fig: measureearly}. Thin lines here represent photonic qubits, thick lines represent feed-forward operations, boxes labelled $X$, $Y$, $Z$, and $H$ represent single qubit $X$, $Y$, $Z$ rotations, and Hadamard gates respectively, and boxes labelled $M_X$, $M_Y$, and $M_Z$ represent measurement in the $X$, $Y$, and $Z$ bases, respectively.

First, we show some results regarding re-ordering of single qubit measurements and rotations. In the left side of Fig.~\ref{fig: measureearly}(a), the unitary operation $U$ on qubit c is conditioned on the result of an $X$ or $Y$ basis measurement on qubit b (that is determined beforehand). In addition, there is a conditional operation $Z^i$ on the qubit b which depends on a feed-forward signal from a different part of the circuit, which in this case is the result of measurement $M_A$ on qubit a. The application of a $Z$ gate before $X$ or $Y$ measurement simply has the effect of flipping the result of the measurement. Hence, the measurement $M_{X}$ (resp. $M_{Y}$) can be performed before $M_A$ and the feed-forward result of $M_A$ can simply be used to flip the result of $M_{X}$ (resp. $M_{Y}$) as shown on the right side of Fig.~\ref{fig: measureearly}(a). The system in Fig. \ref{fig: measureearly}(b) is identical to the system in Fig.~\ref{fig: measureearly}(a) except for the fact that measurement in the $X$ (resp. $Y$) basis is replaced by measurement in the $Z$ basis. Since application of a $Z$ rotation does not influence the outcome of the $Z$ measurement, the $Z$ gate and the associated feed-forward can be removed entirely. In Fig.~\ref{fig: measureearly}(c), we depict that a Hadamard gate followed by an $X$ basis measurement is equivalent to a $Z$ basis measurement, a Hadamard gate followed by a $Z$ basis measurement is equivalent to an $X$ basis measurement, and a Hadamard gate followed by a $Y$ basis measurement is equivalent to a $Y$ basis measurement with the result flipped.

We now use these results to show how measurements can be pushed earlier in the cluster creation process at the major nodes. The left side of Fig.~\ref{fig: measureearly}(d) shows the system with measurements applied after the fusion operations. Single photons that are sent through GHZ factories to create $3$-photon GHZ states, which are then fused using Bell measurements using ancilla photons. The surviving photons require some Hadamard and conditional $Z$ rotations as part of the controlled-phase and parity-projection operations~\cite{2015.PRX.Li-Benjamin.RescostFTLOQC}. Finally, some of the surviving photons require $X$ and $Y$ basis measurements, the results of which are fed forward to photons in the final ``clique" cluster. As shown in Fig.~\ref{fig: measureearly}(a), (b) and (c), measurements in the Pauli basis can be pushed in front of Hadamard and conditional $Z$ rotations by simply moving to a different Pauli basis or flipping the result of the measurement result. Hence, the system is equivalent to the right side of Fig.~\ref{fig: measureearly}(d) in which single qubit Pauli measurements are applied before the fusion operation.

%We welcome any suggestions on this paper. You may contact us directly or submit comments at \url{http://goo.gl/forms/4YA8xRBj9s} (you may submit comments anonymously).

\bibliography{Quantum_Repeater,Error_Correction}

\end{document}